\documentstyle[12pt,titlepage,epsfig,axodraw]{article}

\def\_{\rule{.3em}{.15ex}} 

\newcommand{\scs}{\scriptscriptstyle}
\newcommand{\be}{\begin{equation}}
\newcommand{\ee}{\end{equation}}
\newcommand{\bea}{\begin{eqnarray}}
\newcommand{\eea}{\end{eqnarray}}
\newcommand{\f}{\frac}
\def\slash#1{\setbox0=\hbox{$#1$}#1\hskip-\wd0\dimen0=5pt\advance
       \dimen0 by-\ht0\advance\dimen0 by\dp0\lower0.5\dimen0\hbox
         to\wd0{\hss\sl/\/\hss}}
\def\simleq{\stackrel{<}{\scs \sim}}

\def\mathrm#1{{\rm #1}}
\def\ra{\rightarrow}
\def\kk{\mbox{$\bar{K}^0K^0$~}}
\def\bb{\mbox{$\bar{B}^0B^0$~}}
\def\dd{\mbox{$\bar{D}^0D^0$~}}
\def\bbd{\mbox{$\bar{B}_d^0B_d^0$~}}
\def\BSG{\mbox{$Br(B\ra X_s\gamma)$~}}
\def\bsg{\mbox{$b\ra s\gamma$~}}
\def\epsk{\mbox{$\epsilon_K$~}}
\def\dmbd{\mbox{$\Delta m_{B_d}$~}}
\def\TeV{\mathrm{TeV}}
\def\GeV{\mathrm{GeV}}
\begin{document}

\begin{titlepage}

 \begin{flushright}
  {\bf IFT 3/97\\
       hep-ph/9703442\\
       March 1997}
 \end{flushright}

 \begin{center}
  \vspace{2cm}

\setlength {\baselineskip}{0.3in}
  {\bf \Large Supersymmetry and FCNC Effects}
\vspace{2cm} \\
\setlength {\baselineskip}{0.2in}

{\large  Miko{\l}aj Misiak, 
           Stefan Pokorski 
         and Janusz Rosiek } \\

\vspace{1cm}
{\it Institute of Theoretical Physics, Warsaw University,\\ 
Ho\.za 69, 00-681 Warsaw, Poland}

\vspace{1.5cm} 
{\bf Abstract \\} 
\end{center} 
\setlength{\baselineskip}{0.3in} 

We consider Flavour Changing Neutral Current processes in the
framework of the supersymmetric extension of the Standard Model. FCNC
constraints on the structure of sfermion mass matrices are reviewed.
Furthermore, we analyze supersymmetric contributions to FCNC
transitions which remain in the limit of flavour-conserving sfermion
mass matrices. Implications of the FCNC constraints on the structure
of sfermion mass matrices for SUSY breaking and sfermion mass
generation are discussed. We conclude that the supersymmetric flavour
problem is intriguing but perhaps not as severe as it is commonly
believed. \vspace{2cm}

\begin{quote}
{\bf To appear in the Review Volume ``Heavy Flavours II'', eds.
A.J.~Buras and M.~Lindner, Advanced Series on Directions in High
Energy Physics, World Scientific Publishing Co., Singapore.}
\end{quote}

\end{titlepage} 

\setlength{\baselineskip}{0.3in}

\section{Introduction.}

	Gauge invariance, renormalizability and particle content of
the Standard Model imply the absence (in the lepton sector) or strong
suppression (in the quark sector) of the Flavour Changing Neutral
Current (FCNC) transitions. Such transitions in the quark sector are
absent at the tree level. At one-loop, they are suppressed by light
quark masses (when compared to $M_W$) and by small mixing between the
third and the first two generations. The predicted suppression of the
FCNC processes is in beautiful agreement with the presently available
experimental data. However, the Standard Model is very likely to be
only an effective ``low energy'' theory which, up to some scale
$\Lambda$, is a good approximation to the deeper (and yet unknown)
theory of fundamental interactions. In such a case, renormalizable
interactions of the Standard Model are in general supplemented by
higher dimensional interaction terms suppressed by some powers of the
scale $\Lambda$. These new interactions depend on the structure of the
more fundamental theory. Their $SU(3) \times SU(2) \times U(1)$
invariance is not sufficient any more to protect the observed strong
suppression of the FCNC processes. Consistency with the data then
require that either the scale $\Lambda$ is huge or dangerous new
interactions are absent because of symmetries of the deeper theory.

	The Minimal Supersymmetric Standard Model (MSSM) contains the
new scale $\Lambda$ which is the scale of soft supersymmetry
breaking. It is expected to be of the order of 1~TeV, so long as
supersymmetry is the solution to the so-called hierarchy problem. That
low scale of new physics together with a fully unconstrained
renormalizable minimal\footnote{
\noindent By minimal extension we mean the following three
assumptions:\\ 
(i) minimal particle content consistent with observed
SM particles and SUSY,\\ 
(ii) $SU(3) \times SU(2) \times U(1)$ gauge
invariance,\\ 
(iii) most general soft (dim $<$ 4) SUSY breaking terms
consistent with $SU(3) \times SU(2) \times U(1)$ invariance.}
supersymmetric extension of the SM would be disastrous for the FCNC
transitions.

	As we shall see in the next section, in the MSSM there are,
broadly speaking, two kinds of new contributions to the FCNC
transitions. First of all, they may originate from flavour mixing in
the sfermion mass matrices \cite{D83}. However, even in the absence of
such genuinely new effects, i.e. assuming that the Kobayashi-Maskawa
matrix is solely responsible for flavour mixing, new contributions
arise from charged Higgs boson and chargino exchanges.

	Given the strong suppression of the FCNC transitions observed in
Nature, it is very interesting to study the resulting upper bounds on
flavour changing elements in the sfermion mass matrices and on the
splitting among their diagonal elements. Although these are free
parameters of the MSSM, ultimately their values have to be obtained
from a theory of soft supersymmetry breaking and/or fermion mass
generation.  Therefore, such bounds may provide important hints
towards such a theory. As we shall see, indeed, the sfermion mass
matrices are strongly constrained both in their flavour diagonal and
off-diagonal elements. The weakest are the constraints on the third
generation sfermion masses.

	Since, at the same time, the third generation sfermions and
chargino are expected to be among the lightest superpartners, also the
following question is of obvious interest: Suppose that flavour mixing
in the sfermion mass matrices is small and can be neglected. The
potential impact of supersymmetry on the FCNC transitions appears then
solely through the KM angles present in flavour changing vertices,
i.e. it is due to the charged Higgs and chargino-squark
contributions. Deviations from the SM would be then most probably
first observed in \kk (\epsk parameter), \bb and \dd mixing, as well
as the \bsg transition.  Furthermore, it is reasonable to assume that
the first two generations of sfermions are heavy and degenerate in
mass, and to study the effects which can be generated by light
chargino, charged Higgs boson and the third generation of
sfermions. As we shall see, such a study has also interesting model
independent aspects.

	We find it useful to organize this text according to the two
questions asked above. In section~\ref{sec:formnot}, we briefly
introduce the necessary part of the MSSM notation. In
section~\ref{sec:ndiag}, limits on flavour violation in the sfermion
mass matrices are discussed. In section~\ref{sec:ckm}, we consider
supersymmetric contributions to the FCNC effects from the
chargino--stop (charged Higgs boson -- top) loops, assuming no new
sources of flavour mixing in the sfermion mass matrices.  Finally, in
section~\ref{sec:rge}, we present a brief discussion of the
implications of FCNC on problems like the pattern of soft
supersymmetry breaking or sfermion mass generation.

\section{Formalism and notation.}
\label{sec:formnot}

	We start with a brief description of the MSSM and with
establishing our notation conventions which are similar to the ones
used in ref.~\cite{R90}. The MSSM matter fields are in the following
representations of the $SU(3) \times SU(2) \times U(1)$ gauge
group:\footnote{ The $U(1)$ charges are given in the $SU(5)$
normalization.}

\vspace{0.2cm}
\begin{tabular}{ccccccc}
$     (1,2,-\f{3}{10}) $&$     
      (1,1, \f{3}{5} ) $&$   
      (3,2, \f{1}{10}) $&$ 
(\bar{3},1, \f{1}{5} ) $&$ 
(\bar{3},1,-\f{2}{5} ) $&$ 
      (1,2,-\f{3}{10}) $&$
      (1,2, \f{3}{10}) $ 
\vspace{0.2cm} \\
$ L^I $&$ 
  E^I $&$ 
  Q^I $&$ 
  D^I $&$ 
  U^I $&$ 
  H^1 $&$ 
  H^2 $
\vspace{0.2cm}\\
$ \Psi_{\scs L}^I $&$ 
  \Psi_{\scs E}^I $&$ 
  \Psi_{\scs Q}^I $&$ 
  \Psi_{\scs D}^I $&$ 
  \Psi_{\scs U}^I $&$ 
  \Psi_{\scs H}^1 $&$ 
  \Psi_{\scs H}^2 $
\vspace{0.2cm}
\end{tabular}

\noindent Capital letters in the second row denote complex scalar
fields. The fields in the third row are left-handed fermions. The
upper index $I=1,2,3$ labels the generations. Lower indices (when
present) will label components of $SU(2)$-doublets. Two
$SU(2)$-doublets can be contracted into an $SU(2)$-singlet, e.g. $H^1
H^2 = - H^1_1 H^2_2 + H^1_2 H^2_1$.

	The supersymmetric part of the MSSM Lagrangian schematically
appears as
\bea \label{eq:Lsusy}
{\cal L}_{\scs SUSY} &=& 
-\f{1}{4} F_G^{a\;\mu\nu} F^a_{G\;\mu\nu} 
+ i \bar{\lambda}^a_G {\slash D}_{ab} \lambda^b_G 
+ (D^{\mu}\phi)^{\dagger} (D_{\mu}\phi)
+ i \bar{\psi} {\slash D} \psi \nonumber\\
&& - \left( \f{ \partial \tilde{W}}{\partial \phi_i} \right)^{\star}
  \left( \f{ \partial \tilde{W}}{\partial \phi_i} \right)
-\f{1}{2} \left( \f{ \partial^2 \tilde{W}}{\partial \phi_i \partial 
\phi_j}
           \psi_i^T C \psi_j \; + \; {\rm h.c.} \right) \nonumber\\
&& -\sqrt{2} g_G \left( \phi^{\dagger} T^a_G \lambda^{a\;T}_G C \psi
					\; + \; {\rm h.c.} \right)
-\f{1}{2} g_G^2 (\phi^{\dagger} T^a_G \phi)(\phi^{\dagger} T^a_G \phi).
\eea
where
\be \label{eq:W}
\tilde{W} = \mu H^1 H^2 + Y_l^{IJ} H^1 L^I E^J 
+ Y_d^{IJ} H^1 Q^I D^J + Y_u^{IJ} H^2 Q^I U^J
\label{eq:superpot}
\ee
In eq.~(\ref{eq:Lsusy}), the index $G$ labels the color, weak isospin
and hypercharge factors in the Standard Model gauge group, and indices
$a$ and $b$ range over adjoint representations of the nonabelian
subgroups. All MSSM scalars are assembled into $\phi$, while matter
fermions and gauginos are respectively contained within the
four-component left handed $\psi$ and $\lambda$ fields. The charge
conjugation matrix is denoted by C.

	Apart from the three gauge coupling constants, the supersymmetric
part of the MSSM Lagrangian depends on the Yukawa coupling matrices
$Y_{l,d,u}$ and on the parameter $\mu$ which multiplies the first term
in eq. (\ref{eq:W}).

	The remaining part of the MSSM Lagrangian consists of the soft
supersymmetry breaking terms: gaugino masses, scalar masses and
trilinear scalar interactions
\bea
{\cal L}_{soft} &=& 
-\f{1}{2} \left(   M_3 \tilde{g}^{a\;T} C \tilde{g}^a
                 + M_2 \tilde{W}^{i\;T} C \tilde{W}^i
                 + M_1 \tilde{B}^T C \tilde{B} \; + \; {\rm h.c.} 
				 \right)\nonumber \\ &&
- M_{H^1}^2 H^{1\;\dagger} H^1 - M_{H^2}^2 H^{2\;\dagger} H^2
- L^{I \dagger} ( M_L^2 )^{IJ} L^J 
- E^{I \dagger} ( M_E^2 )^{IJ} E^J
\nonumber \\ &&
- Q^{I \dagger} ( M_Q^2 )^{IJ} Q^J 
- D^{I \dagger} ( M_D^2 )^{IJ} D^J 
- U^{I \dagger} ( M_U^2 )^{IJ} U^J
\nonumber \\ &&
+ \left(   A_E^{IJ} H^1 L^I E^J 
         + A_D^{IJ} H^1 Q^I D^J 
         + A_U^{IJ} H^2 Q^I U^J 
	 + B \mu H^1 H^2 \; + \; {\rm h.c.} \right). \hspace{0.5cm}
\label{eq:lsoft}
\eea

	Assuming the above form of the soft supersymmetry breaking terms,
we depart from full generality. In principle, ${\cal L}_{soft}$ could
be supplemented by all the bilinear and trilinear scalar interaction
terms present in eq.~(\ref{eq:Lsusy}), but with
coupling constants unrelated to those in eq.~(\ref{eq:W}). Here, we
follow the standard approach and assume absence of such terms. Such an
assumption is consistent with renormalization: So long as these terms
are absent at the tree level, they are not generated via loops to all
orders in perturbation theory.

	In the physically acceptable regions of the parameter space,
vacuum expectation values are developed only by the Higgs scalars
\be
\langle H^1 \rangle = 
\left( \begin{array}{c} \f{v_1}{\sqrt{2}} \\ 0 \end{array} \right) 
\equiv
\left( \begin{array}{c} \f{v \cos \beta}{\sqrt{2}} \\ 0 \end{array} 
\right),
\hspace{2cm}
\langle H^2 \rangle = 
\left( \begin{array}{c} 0 \\ \f{v_2}{\sqrt{2}} \end{array} \right) 
\equiv
\left( \begin{array}{c} 0 \\ \f{v \sin \beta}{\sqrt{2}} \end{array} 
\right).
\ee
The value of $v \simeq 246$~GeV is determined from the $W$-boson mass
in the same way as in the Standard Model.

	Lepton and quark mass eigenstates are obtained from the original
left-handed fermion fields with help of $3 \times 3$ unitary matrices
$V^{E,U,D}_{L,R}$ as follows:
\bea 
\begin{array}{rclcrcl}
\nu &=&  V^E_L \Psi_{{\scs L}1}, 
& \hspace{2cm} &
u   &=&  V^U_L \Psi_{{\scs Q}1} \; + \; V^U_R C \bar{\Psi}_{\scs U}^T,
\vspace{0.2cm}\\ 
e   &=&  V^E_L \Psi_{{\scs L}2} \; + \; V^E_R C \bar{\Psi}_{\scs E}^T, 
& \hspace{2cm} &
d   &=&  V^D_L \Psi_{{\scs Q}2} \; + \; V^D_R C \bar{\Psi}_{\scs D}^T.
\end{array}
\eea
Their diagonal $3 \times 3$ mass matrices read
\bea
m_e = -\f{v \cos \beta}{\sqrt{2}} V^E_R Y_e^T V^{E\;\dagger}_L.
\hspace{0.5cm}
m_u =  \f{v \sin \beta}{\sqrt{2}} V^U_R Y_u^T V^{U\;\dagger}_L,
\hspace{0.5cm}
m_d = -\f{v \cos \beta}{\sqrt{2}} V^D_R Y_d^T V^{D\;\dagger}_L,
\nonumber \eea
As in the Standard Model, the Kobayashi-Maskawa matrix is 
$K = V^U_L V^{D\;\dagger}_L$.

	Diagonalization of the scalar mass matrices usually proceeds in
two steps. First, the squarks and sleptons are rotated ``parallel'' to
their fermionic superpartners
\bea
\tilde{N}^0 = V^E_L L_1,
\hspace{0.5cm}
\tilde{L}^0 = \left( \begin{array}{c} V^E_L L_2 \\ V^E_R E^{\star} 
\end{array} 
\right),
\hspace{0.5cm}
\tilde{U}^0 = \left( \begin{array}{c} V^U_L Q_1 \\ V^U_R U^{\star} 
\end{array} 
\right),
\hspace{0.5cm}
\tilde{D}^0 = \left( \begin{array}{c} V^D_L Q_2 \\ V^D_R D^{\star} 
\end{array} 
\right).
\label{eq:superkm}
\eea
The fields in the l.h.s of the above equations form the so-called
``super-KM'' basis in the space of MSSM scalars. These fields may be
often more convenient to work with, even though they are not mass
eigenstates. Their mass matrices have the following form:
\bea
{\cal M}^2_{\tilde{U}} &=&
\left( \begin{array}{cc}
(M^2_{\tilde{U}})_{LL} + m_u^2
- \frac{\cos 2\beta}{6}(M_Z^2 - 4M_W^2)\hat{\mbox{\large 1}} &
(M^2_{\tilde{U}})_{LR} 
- \cot\beta \mu m_u\\
(M^2_{\tilde{U}})_{LR}^{\dagger} 
- \cot\beta \mu^{\star}  m_u& 
(M^2_{\tilde{U}})_{RR} + m_u^2
+\frac{2\cos 2\beta}{3} M_Z^2\sin^2\theta_W\hat{\mbox{\large 1}} \\ 
\end{array}\right),\nonumber\\ \nonumber\\
{\cal M}^2_{\tilde{D}} &=& 
\left( \begin{array}{cc}
(M^2_{\tilde{D}})_{LL} + m_d^2
- \frac{\cos 2\beta}{6}(M_Z^2 + 2M_W^2)\hat{\mbox{\large 1}}  &
(M^2_{\tilde{D}})_{LR} 
- \tan\beta \mu m_d\\
(M^2_{\tilde{D}})_{LR}^{\dagger} 
- \tan\beta \mu^{\star}  m_d& 
(M^2_{\tilde{D}})_{RR} + m_d^2 
-\frac{\cos 2\beta}{3} M_Z^2\sin^2\theta_W\hat{\mbox{\large 1}} \\ 
\end{array}\right), \nonumber\\ \nonumber\\
{\cal M}^2_{\tilde{L}} &=&  
\left( \begin{array}{cc}
(M^2_{\tilde{L}})_{LL}  + m_l^2
+ \frac{\cos 2\beta}{2}(M_Z^2 - 2M_W^2)\hat{\mbox{\large 1}}  &
(M^2_{\tilde{L}})_{LR} 
- \tan\beta \mu m_l\\
(M^2_{\tilde{L}})_{LR}^{\dagger} 
- \tan\beta \mu^{\star} m_l& 
(M^2_{\tilde{L}})_{RR} + m_l^2 
-\cos 2\beta M_Z^2\sin^2\theta_W\hat{\mbox{\large 1}} \ 
\end{array}\right),\nonumber\\ \nonumber\\
{\cal M}^2_{\tilde{N}}& =&
V^E_L M^2_E V_L^{E\dagger} + 
\frac{\cos 2\beta}{2} M_Z^2\hat{\mbox{\large 1}},
\label{eq:sfmass}
\eea

\noindent where $\theta_W$ is the Weinberg angle, $\hat{\mbox{\large
1}}$ stands for the $3 \times 3$ unit matrix, and the flavour-changing
entries are contained in

\bea
\begin{array}{ccc}
(M^2_{\tilde{U}})_{LL} = V_L^U M^2_Q V_L^{U\dagger}
\hspace{0.5cm}&
(M^2_{\tilde{U}})_{RR} = V_R^U M^{2T}_U V_R^{U\dagger}
\hspace{0.5cm}&
(M^2_{\tilde{U}})_{LR} = 
-\f{v \sin \beta}{\sqrt{2}} V_L^U A_U^{\star} V_R^{U\dagger}
\vspace{0.2cm}\\
(M^2_{\tilde{D}})_{LL} = V_L^D M^2_Q V_L^{D\dagger}
\hspace{0.5cm}&
(M^2_{\tilde{D}})_{RR} = V_R^D M^{2T}_D V_R^{D\dagger}
\hspace{0.5cm}&
(M^2_{\tilde{D}})_{LR} = 
 \f{v \cos \beta}{\sqrt{2}} V_L^D A_D^{\star} V_R^{D\dagger}
\vspace{0.2cm}\\
(M^2_{\tilde{L}})_{LL} = V_L^E M^2_L V_L^{E\dagger}
\hspace{0.5cm}&
(M^2_{\tilde{L}})_{RR} = V_R^E M^{2T}_E V_R^{E\dagger}
\hspace{0.5cm}&
(M^2_{\tilde{L}})_{LR} = 
 \f{v \cos \beta}{\sqrt{2}} V_L^E A_E^{\star} V_R^{E\dagger}.
\end{array} 
\eea

	It often happens that certain FCNC processes are sensitive to
particular entries in the above nine matrices. For
$(M^2_{\tilde{U}})_{LL}$, we will denote these entries as follows:

\be (M^2_{\tilde{U}})_{LL} = \left(
\begin{array}{ccc}
  (m^2_{U1})_{LL}    & (\Delta_U^{12})_{LL} & (\Delta_U^{13})_{LL} 
\vspace{0.2cm} \\
(\Delta_U^{21})_{LL} &    (m^2_{U2})_{LL}   & (\Delta_U^{23})_{LL} 
\vspace{0.2cm} \\
(\Delta_U^{31})_{LL} & (\Delta_U^{32})_{LL} &   (m^2_{U3})_{LL}  
\end{array}\right),
\ee 
and analogously for all the other matrices. (Of course
$\Delta^{IJ}_{LL} = \Delta^{JI\star}_{LL}$ and $\Delta^{IJ}_{RR} =
\Delta^{JI\star}_{RR}$, but no such relation holds for $\Delta_{LR}$.)
Experimental constraints will be given on the flavour-changing mass
insertions normalized to a geometric average of the diagonal entries,
e.g.
\be 
(\delta_U^{IJ})_{LR} = 
\f{(\Delta_U^{IJ})_{LR}}{ (m_{UI})_{LL}\;(m_{UJ})_{RR}}.
\ee

	Two remarks are in order here. First, let us suppose that we
have a theory of fermion and sfermion masses which are fixed in some
electroweak basis. We see then, that all four rotations $V_L^U, V_L^D,
V_R^U$ and $V_R^D$ (not just $K$) become partly ``observables''
through the sfermion mass matrices\footnote{
except for ``singular'' cases when the sfermion mass matrices are
diagonal and degenerate or completely aligned with squared Yukawa
coupling matrices in the weak eigenstate basis.}.

	Secondly, one should remember that the matrix
$M^2_{\tilde{Q}}$ is common to the up and down sectors, because of the
$SU(2)$ gauge invariance. Therefore
\be
(M^2_{\tilde{U}})_{LL} = K (M^2_{\tilde{D}})_{LL} K^{\dagger}
\label{eq:udcorr}
\ee
This means that it is impossible to set all the $(\delta^{IJ})_{LL}$
to zero simultaneously, unless $M_Q^2 \sim \hat{\mbox{\large 1}}$, which
implies $(M^2_{\tilde{U}})_{LL} = (M^2_{\tilde{D}})_{LL} \sim 
\hat{\mbox{\large 1}}$.

	Matrices ${\cal M}^2_{\tilde{U}}$ and ${\cal M}^2_{\tilde{D}}$ can
be diagonalized by two additional $6\times 6$ unitary matrices $Z_U$
and $Z_D$, respectively
\bea 
\left({\cal M}^2_{\tilde{U}}\right)^{diag} = 
Z_U^{\dagger} {\cal M}^2_{\tilde{U}} Z_U
\label{eq:zudef}
\\ 
\left({\cal M}^2_{\tilde{D}}\right)^{diag} = 
Z_D^T {\cal M}^2_{\tilde{D}} Z_D^{\star} 
\eea
Of course, if all $\delta^{IJ}$ were zero (i.e. if there was no
flavour mixing in the ``super-KM'' basis), then the matrices $Z_U$ and
$Z_D$ would preserve flavour. Possible off-diagonal entries in these
matrices would then correspond to left-right mixing, i.e. mixing
between superpartners of left- and right-handed quarks having the same
flavour.

	Flavour changing interactions of physical sfermions (mass
eigenstates) depend on the rotations $Z$'s as well as on the mixing
between gauginos and higgsinos. The physical Dirac chargino and
Majorana neutralino eigenstates are linear combinations of left-handed
Winos, Binos and Higgsinos
\bea
\chi^- = (Z^-)^{\dagger} \left( \begin{array}{c} 
\tilde{W}^- \\ (\Psi^1_H)_2 \end{array} \right)
+ (Z^+)^T \left( \begin{array}{c} 
C \overline{\tilde{W}^+}^T \\ C \overline{(\Psi^2_H)}_1^T 
\end{array} \right)\\
\chi^0 = Z_N^{\dagger} \left( \begin{array}{c}
\tilde{B} \\ \tilde{W}_3 \\ (\Psi^1_H)_1 \vspace{0.2cm} \\ (\Psi^2_H)_2 
\end{array} \right) + Z_N^T \left( \begin{array}{c}
C \overline{\tilde{B}}^T \\ C \overline{\tilde{W}}_3^T \\ 
C \overline{(\Psi^1_H)}_1^T \vspace{0.2cm} \\ C 
\overline{(\Psi^2_H)}_2^T 
\end{array} \right).
\eea
The unitary transformations $Z^+$, $Z^-$ and $Z_N$ diagonalize mass
matrices of these fields
\be
{\cal M}_{\chi^{\pm}} = (Z^-)^T \left( \begin{array}{cc}
M_2 & \sqrt{2} M_W \sin \beta \\
\sqrt{2} M_W \cos \beta & \mu \end{array} \right) Z^+
\ee
and
\be
{\cal M}_{\chi^0} = Z_N^T \left( \begin{array}{cccc}
M_1 & 0 & -M_Z \sin \theta_W \cos \beta &  M_Z \sin \theta_W \sin 
\beta \\
0 & M_2 &  M_Z \cos \theta_W \cos \beta & -M_Z \cos \theta_W \sin 
\beta \\
-M_Z \sin \theta_W \cos \beta &  M_Z \cos \theta_W \cos \beta & 0 & 
-\mu \\
M_Z \sin \theta_W \sin \beta & -M_Z \cos \theta_W \sin \beta & -\mu & 0 
\end{array} \right) Z_N.
\ee

\begin{figure}[htbp]
\begin{tabular}{ll}
\begin{picture}(115,60)(0,0)
\DashArrowLine(50,10)(10,10){6}
\Vertex(50,10){2}
\ArrowLine(50,10)(90,10)
\ArrowLine(50,50)(50,10)
\Text(0,10)[]{$U_i$}
\Text(110,10)[]{$(\chi_j^-)^C$}
\Text(50,60)[]{$d^I$}
\end{picture}&
\raisebox{10\unitlength}{\begin{tabular}{l}
$ i\left[\left(-g_2
Z_{U}^{Ji*} Z_{1j}^+ + Y_u^{J} Z_{U}^{(J+3)i*} Z_{2j}^+\right) P_L 
- Y_d^{I} Z_{U}^{Ji*} Z_{2j}^{-*} P_R \right] K^{JI}$\\
\end{tabular}}\\
\begin{picture}(115,60)(0,0)
\DashArrowLine(50,10)(10,10){6}
\Vertex(50,10){2}
\ArrowLine(50,10)(90,10)
\ArrowLine(50,50)(50,10)
\Text(0,10)[]{$D_i$}
\Text(100,10)[]{$\chi_j^-$}
\Text(50,60)[]{$u^J$}
\end{picture}&
\raisebox{10\unitlength}{\begin{tabular}{l}
$ i\left[-\left(g_2
Z_{D}^{Ii} Z_{1j}^+ + Y_d^{I} Z_{D}^{(I+3)i} Z_{2j}^-\right) P_L 
+ Y_u^{J} Z_{D}^{Ii} Z_{2j}^{+*} P_R \right] K^{JI\star}$\\
\end{tabular}}\\
\end{tabular}
\caption{Chargino-quark-squark vertices.}
\label{fig:ccurrent}
\end{figure}
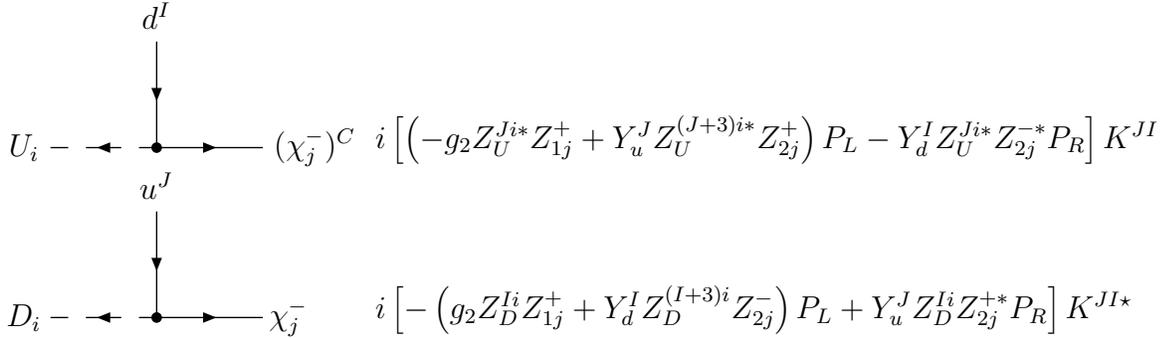

	The most relevant flavour changing vertices for our further
discussion are the ones in which both quarks and squarks are
present. There are three types of such vertices: $f\tilde{f}\chi^-$,
$f\tilde{f}\chi^0$ and $f\tilde{f}\tilde{g}$. They are presented in
Figs.~\ref{fig:ccurrent}-\ref{fig:gcurrent}.

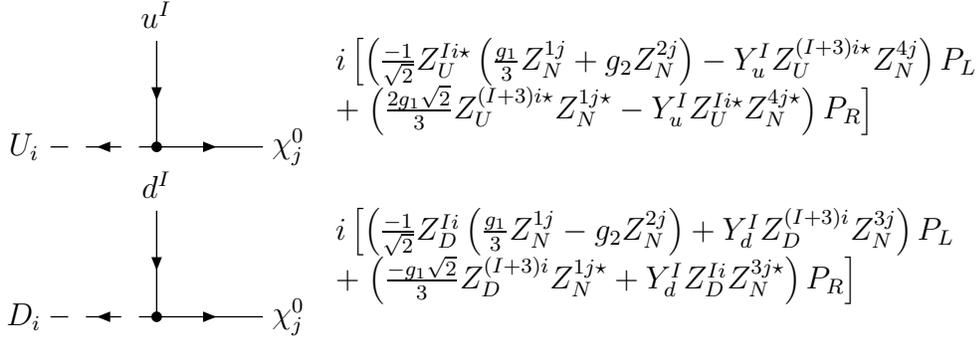
\begin{figure}[htbp]
\begin{tabular}{ll}
\begin{picture}(100,60)(0,0)
\DashArrowLine(50,10)(10,10){6}
\Vertex(50,10){2}
\ArrowLine(50,10)(90,10)
\ArrowLine(50,50)(50,10)
\Text(0,10)[]{$U_i$}
\Text(100,10)[]{$\chi_j^0$}
\Text(50,60)[]{$u^I$}
\end{picture}&
\raisebox{30\unitlength}{\begin{tabular}{l}
$i\left[\left({-1 \over \sqrt{2}} Z_{U}^{Ii\star}\left(\frac{g_1}{3}Z_N^{1j}
+g_2 Z_N^{2j}\right) - Y_u^{I} Z_{U}^{(I+3)i\star} Z_N^{4j}\right) P_L\right.$\\
$+ \left.\left({2g_1 \sqrt{2} \over 3}Z_U^{(I+3)i\star}Z_N^{1j\star}
- Y_u^{I} Z_{U}^{Ii\star} Z_N^{4j\star}\right) P_R \right]$\\
\end{tabular}}\\
\begin{picture}(100,60)(0,0)
\DashArrowLine(50,10)(10,10){6}
\Vertex(50,10){2}
\ArrowLine(50,10)(90,10)
\ArrowLine(50,50)(50,10)
\Text(0,10)[]{$D_i$}
\Text(100,10)[]{$\chi_j^0$}
\Text(50,60)[]{$d^I$}
\end{picture}&
\raisebox{30\unitlength}{\begin{tabular}{l}
$i\left[\left({-1 \over \sqrt{2}} Z_{D}^{Ii}\left(\frac{g_1}{3}Z_N^{1j}
-g_2 Z_N^{2j}\right)+ Y_d^{I} Z_{D}^{(I+3)i} Z_N^{3j}\right) P_L\right.$\\
$+ \left.\left({-g_1 \sqrt{2} \over 3}Z_D^{(I+3)i}Z_N^{1j\star}
+ Y_d^{I} Z_{D}^{Ii} Z_N^{3j\star}\right) P_R \right]$\\
\end{tabular}}\\
\end{tabular}
\caption{Neutralino-quark-squark vertices.}
\label{fig:ncurrent}
\end{figure}
\input gcurrent.axo

	As an example of how these vertices enter FCNC amplitudes, let
us list supersymmetric contributions to the \kk mixing. All the MSSM
diagrams are shown in Fig.~\ref{fig:kkmix}.  In addition to the
Standard Model ($W-q$) box diagrams, we have the charged Higgs,
chargino, neutralino and gluino exchanges. In these diagrams, all the
particles are mass eigenstates, and the vertices depend on the
rotations $Z_U$ and $Z_D$. 
\input kkmix.axo

	It is often convenient to work in the super-KM basis, provided
an approximation to the first nonvanishing order in $\Delta^{IJ}$ is
sufficient. In such a basis, for instance, the diagram (B) in
Fig.~\ref{fig:kkmix} is replaced by its expansion shown in
Fig.~\ref{fig:ckkmix}. In Fig.~\ref{fig:ckkmix}, the fields
$\tilde{U}^0=(\tilde{U}^0_L,\tilde{U}^0_R)$ and
$\tilde{C}^0=(\tilde{C}^0_L,\tilde{C}^0_R)$ are up-squark fields in
the super-KM basis, and the vertices are given by the formulae in
Figs.~\ref{fig:ccurrent}-\ref{fig:ncurrent}, with flavour-preserving
$Z_U$ and $Z_D$. Similarly, in the super-KM basis, the diagram (C) in
Fig.~\ref{fig:kkmix} is replaced by its expansion plotted in
Fig.~\ref{fig:gkkmix}.
\input ckkmix.axo
\input gkkmix.axo

	Two important remarks at this point are the following:
As we have already said in the introduction, supersymmetric
contributions to the FCNC transitions arise even when fermion
and sfermion mass matrices are simultaneously flavour-conserving
($\Delta^{IJ}=0$).  These contributions originate from the
charged Higgs and chargino exchange diagrams in
Fig.~\ref{fig:kkmix}, with the KM angles in the vertices.
Similar diagrams contribute to the \bb and \dd mixing.

	Secondly, in a general case with $\Delta^{IJ}\neq 0$, the
dependence on particular mass insertions enters into various processes
in a correlated way. Correlated vertices are shown in
Fig.~\ref{fig:delcorr},
\input delcorr.axo
with $(M^2_{\tilde{U}})_{LL} = K (M^2_{\tilde{D}})_{LL} K^{\dagger}$
(see eq.~(\ref{eq:udcorr})).  (However, the right-handed elements are
uncorrelated). These formulae relate, for example, chargino and
gluino/neutralino contributions to neutral meson mixing.  In addition,
the same elements ($\Delta_{LL}$, $\Delta_{RR}$ and $\Delta_{LR}$) may
enter various processes, e.g. as illustrated in Fig.~\ref{fig:idcorr}.
\input idcorr.axo
These correlations have to be taken into account in a complete and
systematic study of FCNC transitions in the MSSM.

	Having presented the necessary formalism, we proceed to discussing
bounds on the sfermion mass matrices obtained from the
experimentally observed strong suppression of the FCNC effects.

\section{Bounds on sfermion masses from FCNC processes}
\label{sec:ndiag}

	Strong experimental suppression of the FCNC transitions puts
severe upper bounds on various entries in the sfermion mass matrices
of eq.~(\ref{eq:sfmass}) at low energy. Such bounds are of crucial
interest for the theory (as yet unknown) of soft supersymmetry
breaking. As we have already mentioned, a systematic discussion of
such bounds should include all potential contributions and
correlations among them. However, in the first approximation, one can
neglect all but the gluino (photino) exchange contributions to the
FCNC transitions in the quark (lepton) sector.  Order-of-magnitude
bounds on the off-diagonal entries in the squark mass matrices are
then obtained under the assumption that these contributions saturate
the experimental results \cite{HKT94,GGMS96}. Bounds on splittings
between diagonal elements of $(M^2_{\tilde{U}})_{LL}$ and
$(M^2_{\tilde{D}})_{LL}$ can then be obtained with use of
eq. (\ref{eq:udcorr}). Such bounds have a virtue of being relatively
parameter independent.

	In the next step, it is also interesting to discuss how these
bounds can be modified in a complete analysis, with all contributions
and interference between them included. We shall see that
cancellations can indeed occur and weaken the limits.  However, large
cancellations affecting their order of magnitude would require certain
fine-tuning of the MSSM parameters. Thus, we may conclude that the
bounds obtained from the gluino (photino) contributions indeed reflect
the acceptable structure of the mass matrices, at least up to an order
of magnitude.

	The most up-to-date set of bounds on the flavour off-diagonal
entries in the sfermion mass matrices is given in ref.~\cite{GGMS96}
(for earlier results see \cite{HKT94,GGMS96} and references
therein). We shall not repeat details of those analyses here. Very
briefly, the bounds on $\delta_D^{12}$, $\delta_D^{13}$ and
$\delta_U^{12}$ are obtained from gluino contributions to the \kk,
\bbd and \dd mixing, respectively. The contributions are given by box
diagrams which are proportional to biproducts of $\delta$'s of
different chiralities. The bounds are obtained by assuming that each
term of the expansion in the biproducts od $\delta$'s saturates by
itself the experimental results.

	In the analysis of neutral meson mixing, one usually
introduces an effective hamiltonian built out of flavour changing
four-quark operators. The operators arising in the SM are built out of
left-handed quark fields only. They arise in the MSSM, too. However,
the MSSM interactions can generate in addition a whole set of extra
operators containing quarks of both chiralities. In the case of
SM-like operators, the effect of supersymmetry is seen in a
modification of their Wilson coefficients. Matrix elements of these
operators between neutral meson states are the same as in the SM.
Their values are estimated from lattice calculations and parametrized
by quantities denoted by e.g. $B_K$ or $B_{B_d}$. However, no lattice
results are available for the extra operators. One has to rely on
rough PCAC estimates only (see refs.~\cite{HKT94,GGMS96} for more
details). However, this is still a correct approach, so long as only
order-of-magnitude bounds on $\delta^{IJ}$ are being estimated.

	Due to the Appelquist-Carazzone decoupling, bounds on
$\delta^{IJ}$ become weaker when masses of squarks and gluinos
increase. When all the squark masses $m_{\tilde{q}}$ are close in
size, we can parametrize this suppression by 
$m_{max} = {\rm max}( m_{\tilde{q}},m_{\tilde{g}})$. 
Neutral meson mixing gives us bounds on $\delta^{IJ}/m_{max}$. On the
other hand, the limits do not depend strongly on the ratio
\be \label{eq:rratio}
r_{\tilde{q}\tilde{g}} = 
\f{{\rm min}( m_{\tilde{q}},m_{\tilde{g}})}
  {{\rm max}( m_{\tilde{q}},m_{\tilde{g}})}.
\ee
Even changing $r_{\tilde{q}\tilde{g}}$ in its whole domain $[0,1]$
(for both $m_{\tilde{q}} > m_{\tilde{g}}$ and $m_{\tilde{q}} <
m_{\tilde{g}}$) results in changing bounds on e.g.
$(\delta_D^{12})_{LL}/m_{max}$ by less than an order of
magnitude.\footnote{
This is not true inside a small region $1.4 <
m_{\tilde{g}}/m_{\tilde{q}} < 1.7$ where accidental cancellations
occur. In this region, there is a point where gluino-squark
contributions to neutral meson mixing give no limit on $\delta_{LL}$
alone.\label{foot:finetune}}

	For $r_{\tilde{q}\tilde{g}} = 1$ the bounds on flavour changing
entries in the squark mass matrices can be summarized as follows:
\bea
|(\delta_D)_{LL}|,  |(\delta_D)_{RR}|   & \simleq &
\left( \begin{array}{c|c|c} 
~~~~~~~~~& 
0.08 \f{m_{max}}{1\;{\rm TeV}} &
0.2  \f{m_{max}}{1\;{\rm TeV}}\\
\hline
&~~~~~~~~~~~~~~& 30 (\f{m_{max}}{1\;{\rm TeV}})^2\\
\hline
&& ~~~~~~~~~~~~~
\end{array} \right),\vspace{1cm}  \hspace{2cm} \label{eq:dDLL.dDRR}\\
|(\delta_D)_{LR}| & \simleq &
\left( \begin{array}{c|c|c} 
~~~~~~~~~&  
0.009 \f{m_{max}}{1\;{\rm TeV}} & 0.07  \f{m_{max}}{1\;{\rm TeV}}\\
\hline
0.009 \f{m_{max}}{1\;{\rm TeV}} 
&~~~~~~~~~~~~~~&
0.03 \f{m_{max}}{1\;{\rm TeV}}\\
\hline
0.07 \f{m_{max}}{1\;{\rm TeV}} & 0.03  \f{m_{max}}{1\;{\rm TeV}} &
~~~~~~~~\end{array} \right),\vspace{1cm} \hspace{2cm} \label{eq:dDLR}\\
|(\delta_U)_{LL}|   & \simleq &
\left( \begin{array}{c|c|c} 
~~~~~~~~~& 
0.2 \f{m_{max}}{1\;{\rm TeV}} &
   {\cal O}( {\rm max}[ (\delta_D^{13})_{LL},
                        (\delta_U^{12})_{LL} \f{K^{13}}{K^{12}}])\\
\hline
&~~~~~~~~~~~~~~& 
{\cal O}( {\rm max}[ (\delta_D^{23})_{LL},
                        (\delta_U^{12})_{LL} \f{K^{23}}{K^{12}}])\\
\hline
&& ~~~~~~~~~~~~~\end{array} \right),\vspace{1cm} \hspace{2cm} 
\label{eq:dULL}\\
|(\delta_U)_{RR}|   & \simleq &
\left( \begin{array}{c|c|c} 
~~~~~~~~~&  0.2 \f{m_{max}}{1\;{\rm TeV}} & ~~~?~~~\\
\hline
&~~~~~~~~~~~~~~& ?\\
\hline
&& ~~~~~~~~~~~~~\end{array} \right),\vspace{1cm} \hspace{2cm} 
\label{eq:dURR}\\
|(\delta_U)_{LR}| & \simleq &
\left( \begin{array}{c|c|c} 
~~~~~~~~~&  
0.2 \f{m_{max}}{1\;{\rm TeV}} & ~~~\star~~~\\
\hline
0.2 \f{m_{max}}{1\;{\rm TeV}} &~~~~~~~~~~~~~~& \star\\
\hline
? & ? & 
~~~~~~~~\end{array} \right).\vspace{1cm} \hspace{2cm}\label{eq:dULR}
\eea
The off-diagonal entries which can be obtained from hermiticity have
been left empty. Question marks denote unconstrained entries which
would require experimental information on rare top quark decays. Stars
denote the entries which are unconstrained by gluino exchanges but
receive bounds from chargino-squark loops (see ``note added''). Bounds
on the (23) and (32) entries of $\delta_D$ matrices were obtained from
\bsg decay. Estimates for (13) and (23) entries in $(\delta_U)_{LL}$
are found from the relation (\ref{eq:udcorr}). We discuss consequences
of this relation in more detail below.

	Limits on leptonic $\delta_L$ obtained from $l^I\ra l^J\gamma$
decays are as follows:
\bea
|(\delta_L)_{LL}|,  |(\delta_L)_{RR}|   & \simleq &
\left( \begin{array}{c|c|c} 
 ~~~~~~~~~~~~~~~~~~& 
0.2  (\f{m_{max}}{500\;{\rm GeV}})^2 &
700 (\f{m_{max}}{500\;{\rm GeV}})^2 \\
\hline
&~~~~~~~~~~~~~~~~~~& 100 (\f{m_{max}}{500\;{\rm GeV}})^2\\
\hline
&& \end{array} \right),\vspace{1cm}  \hspace{2cm} \label{eq:dLLL.dLRR}\\
|(\delta_L)_{LR}| & \simleq &
\left( \begin{array}{c|c|c} 
 ~~~~~~~~~~~~~~~~~~&  
1 \times 10^{-5} \f{m_{max}}{500\;{\rm GeV}} & 
0.5 \f{m_{max}}{500\;{\rm GeV}}\\
\hline
1 \times 10^{-5} \f{m_{max}}{500\;{\rm GeV}}
&~~~~~~~~~~~~~~~~~~& 0.1 \f{m_{max}}{500\;{\rm GeV}} \\
\hline
0.5 \f{m_{max}}{500\;{\rm GeV}} & 0.1  \f{m_{max}}{500\;{\rm GeV}} 
&~~~~~~~~~~~~~~~~
\end{array} \right).\vspace{1cm}  \hspace{2cm} \label{eq:dLLR}
\eea
In this case, $m_{max}$ stands for ${\rm max}(
m_{\tilde{l}},m_{\tilde{\gamma}})$. The presented numerical bounds
correspond to equal slepton and photino masses, i.e. to
$r_{\tilde{l}\tilde{\gamma}} = 1$. The ratio
$r_{\tilde{l}\tilde{\gamma}}$ is defined analogously to
eq.~(\ref{eq:rratio}). The bounds do not depend strongly on this
ratio, similarly to the squark-gluino case.

	It is important to note that the limits on the $\delta_{LR}$
matrices originating from gluino and photino loops are symmetric not
because the matrices themselves are symmetric, but because the
considered amplitudes depend on their off-diagonal entries in a
symmetric manner. Furthermore, we have to mention that we have
identified absolute values of all the entries with their real
parts. This is reasonable, because CP-violating phenomena put bounds
on the imaginary parts which are usually much stronger than bounds on
real parts. In ref. \cite{GGMS96}, one can find explicit bounds on the
imaginary parts of $\delta_U$, $\delta_D$ and $\delta_L$.

	The method applied for finding bounds on $|\delta_U|$ and
$|\delta_D|$ gives us ``independent'' limits on certain products of
these entries. too. For instance
\bea
\sqrt{|(\delta_D^{12})_{LL}(\delta_D^{12})_{RR}|}
& \simleq & 0.006 \; \f{m_{max}}{1\;{\rm TeV}}
\nonumber \\
\sqrt{|(\delta_D^{13})_{LL}(\delta_D^{13})_{RR}|}
& \simleq & 0.04 \; \f{m_{max}}{1\;{\rm TeV}}
\label{eq:sqrtLLRR} \\
\sqrt{|(\delta_U^{12})_{LL}(\delta_U^{12})_{RR}|}
& \simleq & 0.04 \; \f{m_{max}}{1\;{\rm TeV}}. \nonumber 
\eea

	These bounds look more restrictive than the previously given
bounds on $\delta^{IJ}_{LL}$ and $\delta^{IJ}_{RR}$
separately. Actually, the allowed region in the
$Re(\delta^{IJ}_{LL})$--$Re(\delta^{IJ}_{RR})$ plane (for given (IJ))
is bounded by two hyperbolae centered at the origin. The symmetry axes
of these hyperbolae are close to being horizontal or vertical. This is
why the product of the two $\delta$'s is more restricted than each
$\delta$ alone. No strict bound on $\delta$'s exists when fine-tuning
between them is allowed. Barring fine-tuning, one can only conclude
that the expected sizes of $\delta$'s are somewhere between those in
eq. (\ref{eq:sqrtLLRR}) and those in eqs. (\ref{eq:dDLL.dDRR}) and
(\ref{eq:dULL}).

	It is interesting to notice, that box diagrams constrain
$\delta/m_{max}$, while the penguin ones give bounds on
$\delta/m_{max}^2$. An exception from the latter rule are bounds on
$(\delta^{IJ})_{LR}$ from processes receiving important contributions
from dimension 5 effective operators.\footnote{
All such operators can be reduced by equations of motion to the
so-called ``magnetic moment'' operators, like the two we give later in
eqs. (\ref{P7}) and (\ref{P8}).}
Such processes (like \bsg) give us bounds on
$(\delta^{IJ})_{LR}/m_{max}$. Since the limits on squark mass matrices
we have listed above originate from box diagrams and from \bsg, only
one of these constraints scales like $m_{max}^2$.

	The limits we have discussed so far (following
ref.~\cite{GGMS96}) are derived from gluino and photino exchange
contributions to various FCNC processes. It is also interesting to
consider bounds originating from diagrams with chargino exchanges.  As
an example, chargino--(up~squark) contribution to \bsg decay is
discussed in more detail in~\ref{app:bsgct}. As seen
in~\ref{app:bsgct}, chargino diagrams restrict certain linear
combinations of diagonal mass splittings and off-diagonal elements of
$(M^2_U)_{LL}$ (at the leading order in these quantities). Those
linear combinations turn out to be equal to the off-diagonal elements
of $(M^2_D)_{LL}$ only. The relation (\ref{eq:udcorr}) is essential
for making this observation.

	We argue in~\ref{app:bsgct} that the same conclusion holds for
other processes: Chargino, neutralino and gluino contributions to
processes involving down quarks in the initial and final states
(e.g. \bsg decay, \kk and \bbd mixing) are all directly sensitive to
the structure of $(M^2_D)_{LL}$, not $(M^2_U)_{LL}$. Similarly,
processes involving up quarks in the initial and final states (like
\dd mixing) are directly sensitive to the structure of
$(M^2_U)_{LL}$. The constraints on $(\delta_{U,D})_{LL}$ from chargino
loops are expected to be of the same order of magnitude as the gluino
ones.\footnote{
Suppression by electroweak coupling is off-set by relatively smaller
chargino mass, at least when GUT relations between gaugino masses are
assumed.} 
On the other hand, bounds from chargino loops on $(\delta_{U,D})_{LR}$
and $(\delta_{U,D})_{RR}$ (except for $(\delta_U^{13})_{LR}$ and
$(\delta_U^{23})_{LR}$ ) are inefficient due to small Yukawa couplings
of the first two generations.

	Let us now turn to restrictions on diagonal entries of the
matrices $(M^2_U)_{LL}$ and $(M^2_D)_{LL}$. Using again
eq.~(\ref{eq:udcorr}), we can express splitting between these diagonal
entries in terms of the off-diagonal ones. The exact formulae are the
following:
\bea
(m_{UI}^2)_{LL} - (m_{UJ}^2)_{LL} &=& 
 \f{1}{K^{IJ}} [ K (\Delta_D)_{LL} - (\Delta_U)_{LL} K ]^{IJ}
-\f{1}{K^{JJ}} [ K (\Delta_D)_{LL} - (\Delta_U)_{LL} K ]^{JJ},
\nonumber \\ \ \label{eq:splitU} \\
(m_{DI}^2)_{LL} - (m_{DJ}^2)_{LL} &=& 
 \f{1}{K^{IJ}} [ K (\Delta_D)_{LL} - (\Delta_U)_{LL} K ]^{IJ}
-\f{1}{K^{II}} [ K (\Delta_D)_{LL} - (\Delta_U)_{LL} K ]^{II},
\nonumber \\ \ \label{eq:splitD}
\eea
where no summation over the indices $I$ and $J$ is
understood.\footnote{
A relation between $\Delta_U$ and $\Delta_D$ which is independent of
the diagonal entries can be obtained e.g. from the first equation by
adding its $(IJ) = (12), (23)$ and $(31)$ components. This is where
the estimates for $(\delta_U^{13})_{LL}$ and $(\delta_U^{23})_{LL}$ in
eq. (\ref{eq:dULL}) originate from.}
Our previous discussion implies that eqs.~(\ref{eq:splitU}) and
(\ref{eq:splitD}) are the only available source of information
concerning diagonal mass splittings in $(M^2_U)_{LL}$ and
$(M^2_D)_{LL}$, up to ${\cal O}((\delta m^2/m^2)^2)$ effects.

	The presence of $1/K^{IJ}$ in the constraints on mass
splitting in eqs.~(\ref{eq:splitU}) and (\ref{eq:splitD}) makes these
bounds completely inefficient for the third generation of
squarks. Even the bound on the splitting between the first two
generations is rather weak when bounds on $\delta$'s from gluino
exchanges are used. Approximately, it reads
\be \label{eq:numsplit12}
\f{|(m_{U1}^2)_{LL} - (m_{U2}^2)_{LL}|}{m_{\tilde{q}}^2}, \; 
\f{|(m_{D1}^2)_{LL} - (m_{D2}^2)_{LL}|}{m_{\tilde{q}}^2} 
\simleq \f{|(\delta_U^{12})_{LL}| + |(\delta_D^{12})_{LL}|}{K^{12}} 
\simeq 1 \times \f{m_{max}}{1\;{\rm TeV}}
\ee
This bound could become a factor of two lower if the experimental
constraints on \dd mixing improved by the same factor. Furthermore,
nonvanishing $\delta_{RR}$ would improve it (indirectly) as well,
because of correlations between $\delta_{LL}$ and $\delta_{RR}$ (see
eq. (\ref{eq:sqrtLLRR}) and below). However, one should keep in mind
that all the bounds we discuss here are only order-of-magnitude ones.

	We have already mentioned that chargino loops give us direct
constraints only on the off-diagonal elements of
$(M^2_{\tilde{U}})_{LL}$ and $(M^2_{\tilde{D}})_{LL}$. Consequently,
bounds from chargino diagrams on diagonal mass splittings in these
matrices can be derived from eqs.~(\ref{eq:splitU}) and
(\ref{eq:splitD}) only. They are similar to those given in
eq.~(\ref{eq:numsplit12}). On the other hand, direct bounds from
chargino diagrams on diagonal mass splittings in the left-right and
right-right blocks of squark mass matrices are inefficient due to
small Yukawa couplings of the first two generations. This is because
winos couple to left-handed quarks only while higgsino couplings to
the first two generations are very small.

 	Another way of restricting the off-diagonal elements and
diagonal mass splittings in the squark mass matrices is to require
that supersymmetric contributions to FCNC processes do not exceed the
Standard Model ones. This allows to see more easily the relation
between squark mass splittings and the GIM mechanism in the SM. As an
example, let us consider this part of the gluino contribution to
$\Delta m_K$ which is proportional to
$(\delta^{12}_D)_{LL}$. Requiring that it is not larger than the
(QCD-uncorrected) short-distance SM contribution, we find for
$r_{\tilde{q}\tilde{g}} = 1$
\be \label{eq:GIMdLL}
|(\delta^{12}_D)_{LL}| \simleq \sqrt{\f{27}{2}} 
\f{\alpha_2}{\alpha_3} K^{12} \f{m_c m_{\tilde{q}}}{M_W^2},
\ee
which agrees (within a factor of 2) with the bound quoted in
eq. (\ref{eq:dDLL.dDRR}). As usual, we have neglected the imaginary
part of $(\delta^{12}_D)_{LL}$. Inserting the above bound into
eq.~(\ref{eq:numsplit12}), one finds 
\be
|m_{\tilde{q}1} - m_{\tilde{q}2}| \simleq
2 \sqrt{\f{27}{2}} \f{\alpha_2}{\alpha_3} m_c \f{m_{\tilde{q}}^2}{M_W^2}
\simeq 2 m_c \f{m_{\tilde{q}}^2}{M_W^2} \; + \; {\cal O}
((\delta_U^{12})_{LL})
\ee
for both up and down squarks. Thus, if masses of squarks and gluinos
were close to $M_W$ and $(\delta_U^{12})_{LL}$ was negligibly small,
then differences between masses of left squarks of the first two
generations would need to be close to the charm quark mass (or
smaller). This would mean degeneracy by at most a few percent. On the
other hand, if masses of squarks and gluinos were close to 1~TeV (but
$(\delta_U^{12})_{LL}$ was still negligible), the first two left
squark generations could differ in mass by even 50\%. These
restrictions get weaker by about a factor of 2 to 3 when we take into
account nonvanishing $(\delta^{12}_U)_{LL}$ within the bounds allowed
by \dd mixing data (eq. (\ref{eq:dULL})).

	Bounds on off-diagonal elements and diagonal mass splittings in
the squark mass matrices are sensitive to
\begin{figure}[htbp]
\begin{center}
\begin{tabular}{p{0.48\linewidth}p{0.48\linewidth}}
\mbox{\epsfig{file=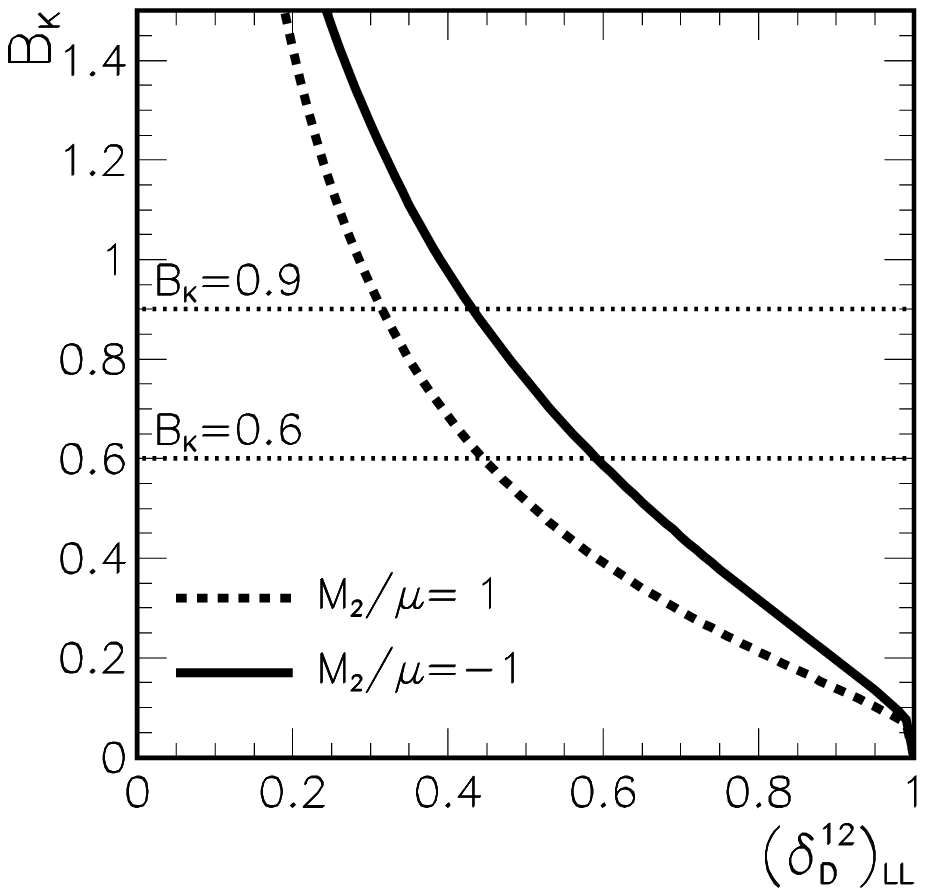,width=\linewidth}}
&
\mbox{\epsfig{file=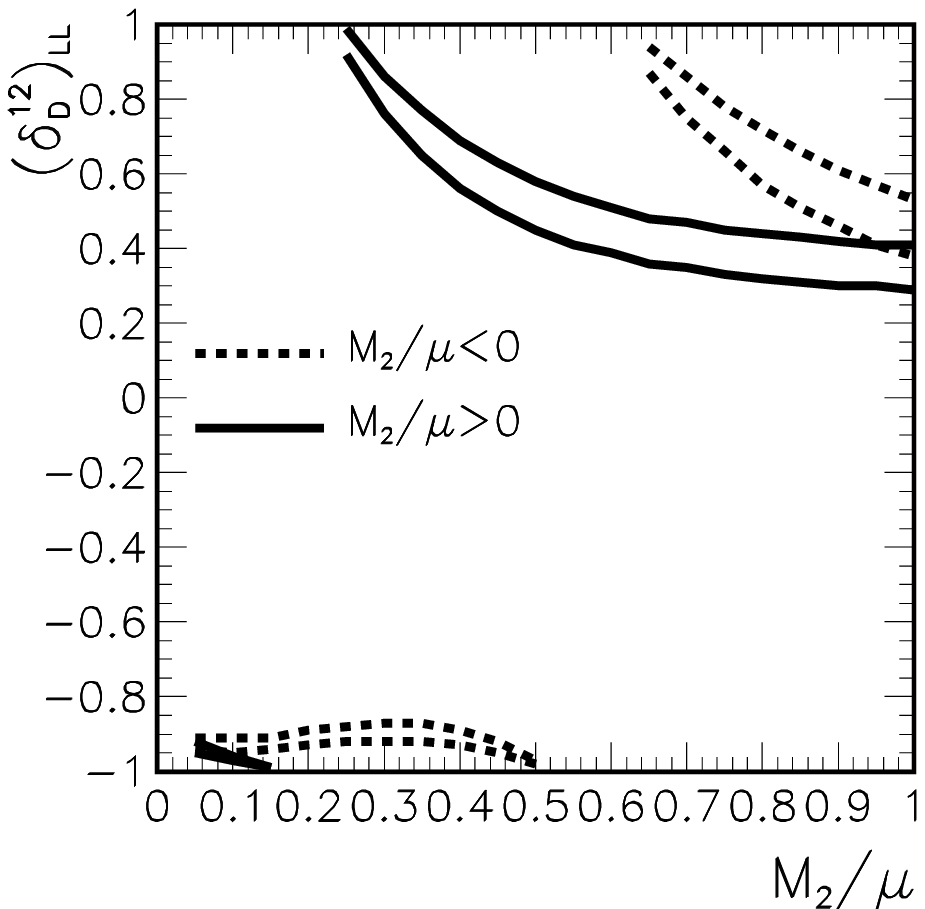,width=\linewidth}}
\\
\caption{Values of $B_K$ necessary to restore the experimental result
for \epsk as a function of $(\delta^{12}_D)_{LL}$ for a chosen set of
SUSY parameters (see the text).
\label{fig:dmixl12}}&
\caption{Upper and lower bounds on $(\delta^{12}_D)_{LL}$ as a
function of $M_2/\mu$ for a chosen set of
SUSY parameters (see the text).
\label{fig:lim12}}\\
\end{tabular}
\end{center}
\end{figure}
interference between chargino and gluino contributions. Sizable
effects in bounds derived from neutral meson mixing may be observed
mainly when the limits on $(\delta_{U,D})_{LL}$ are considered.  In
the LR and RR cases, chargino contributions are suppressed by small
Yukawa couplings whenever both gluinos and charginos contribute
proportionally to the same $\delta$. In
Figs.~\ref{fig:dmixl12},\ref{fig:lim12} we present an example of
bounds on $(\delta^{12}_D)_{LL}$ following from the \epsk measurement
for a chosen set of SUSY parameters and KM phase: $\tan\beta=1.8$,
$m_{\chi_1^{\pm}}=m_{\tilde{T}_R}=100$ GeV,
$m_{gluino}=m_{\tilde{T}_L}=500$ GeV, $m_{H^{\pm}}=1000$ GeV,
$\theta_{LR}=0$, $\sin\delta^{KM}=0.7$. In Fig.~\ref{fig:dmixl12}, we
plot values of the hadronic matrix element parameter $B_K$ necessary
to restore the experimental result $\epsk=(2.26\pm 0.02)\,10^{-3}$ for
given value of $(\delta^{12}_D)_{LL}$ and compare them with
theoretical estimates $0.6\leq B_K\leq 0.9$~\cite{B96}. Acceptable
range for $B_K$ is denoted by dotted horizontal lines which determine
the bounds on $(\delta^{12}_D)_{LL}$. As can be seen from
Fig.~\ref{fig:dmixl12}, detailed limits on $(\delta^{12}_D)_{LL}$
depend on the chargino mixing angles determined by $M_2/\mu$. An
example of such a dependence is shown in Fig~\ref{fig:lim12}. Bounds
on $(\delta^{12}_D)_{LL}$ obtained from the condition (see eq.
(\ref{eq:epsk}) in the next section):
\bea
0.6\leq B_K((\delta^{12}_D)_{LL}) = 
{\epsilon_K^{exp}\over \epsilon_K^{theor}(B_K=1,(\delta^{12}_D)_{LL})}
\leq 0.9
\eea
are plotted there as a function of $M_2/\mu$.  We see that when
cancellations between the chargino and gluino contributions occur, the
bounds on $(\delta^{12}_D)_{LL}$ can be weaker even by an order of
magnitude (or more). For some values of $M_2/\mu$ in
fig.~\ref{fig:lim12}, chargino and gluino contributions cancel
exactly, and bounds on $(\delta^{12}_D)_{LL}$ disappear completely.
Thus, if we are ready to accept some degree of fine-tuning, the bounds
on $(\delta_{U,D}^{IJ})_{LL}$ can be significantly weaker than those
given in matrices~(\ref{eq:dDLL.dDRR},\ref{eq:dULL}). However, we
should stress that there is no similar mechanism for weakening the
bounds on $(\delta_{U,D}^{IJ})_{LR}$, $(\delta_{U,D}^{IJ})_{RR}$.
Therefore, it is clear from eq.~(\ref{eq:sqrtLLRR}) that the overall
weakening of the bounds given in~(\ref{eq:dDLL.dDRR},\ref{eq:dULL})
can only be moderate. We may conclude again
that~(\ref{eq:dDLL.dDRR},\ref{eq:dULL}) reflect the expected order of
magnitude of bounds on $\delta$'s, even in the presence of some
fine-tuning.

	In the end of this section, let us make a comment about FCNC
processes other than \bsg and neutral meson mixing. As far as
processes involving down quarks in the initial and final states are
considered, one can expect that bounds roughly similar to those in
eqs. (\ref{eq:dDLL.dDRR})--(\ref{eq:dULR}) can be derived by requiring
that SUSY amplitudes do not exceed SM ones. However, experimental
constraints on FCNC processes other than \kk and \bb mixing as well as
\bsg are usually well above the (short-distance) SM predictions. This
is why \bsg and neutral meson mixing are most restrictive for most
SUSY parameter choices.

	Nevertheless, other processes can be helpful in some limited
domains of the MSSM parameter space. For instance, large SUSY
contributions to $b \to s\;gluon$ can be sometimes obtained without
violating \bsg bounds $\cite{CGG96}$. Certain asymmetries in $b \to s
e^+ e^-$ are essential to determine the sign of \bsg amplitude
\cite{AGM95,CMW96}, which matters in studying the allowed MSSM
parameter space. Last but not least, various CP-violating observables
like $\epsilon'/\epsilon$, electric dipole moments or $K_L \to \pi^0
\nu \bar{\nu}$ decay rate are essential in verifying whether other
than KM phase sources of CP violation occur in the MSSM \cite{GNR97}.

	As far as FCNC processes involving up quarks are considered,
improving experimental bounds on \dd mixing seems more promising than
studying processes like $c \to u \gamma$. Mixing with the third
generation can be restricted only when rare top quark decays become
experimentally accessible. Before this happens, some of the
superpartners may be already discovered.

\section{ FCNC with light superpartners.}
\label{sec:ckm}

	The bounds discussed in section~\ref{sec:ndiag} must be
satisfied in any realistic supersymmetric extension of the SM. It
should be stressed that there are two general ways of achieving
this. The most straightforward solution occurs (see
e.g. ref. \cite{HIER12}) when sfermions of the first two generations
are sufficiently heavy, so that new contributions to the FCNC
processes in $(1,2)$ sector decouple by the Appelquist-Carrazone
theorem~\cite{AC75}. Indeed, the strongest bounds are for (1,2)
sector. Satisfying them for $\delta^{12}\sim{\cal O}(1)$ requires
quite large masses of the first two generations of sfermions
$m_{1,2}\sim{\cal O}(10~\TeV)$. On the other hand, the constraints in
the $(1,3)$ and $(2,3)$ sector are much weaker and satisfying them
with $m_3\sim{\cal O}(M_Z)$, i.e. $\sqrt{m_3 m_{1,2}}\sim{\cal
O}(1~\TeV)$ is easier. As we discuss in the last section, this
possibility is not at all unnatural, and does not ruin the virtues of
supersymmetry as a solution to the hierarchy problem.
 
	Another possibility is that for some deeper theoretical
reasons\footnote{
Some speculative ideas are collected in section~\ref{sec:rge}.}
all the $\delta^{IJ}$ ($I,J=1..3$) are indeed very small at low
energies. In addition, if high degeneracy of the first two sfermion
generations occurs, their masses are bounded from below only by the
present direct search limits. These limits are close to 200~GeV
at present \cite{PDG96}.

	It is interesting to observe that both solutions allow for
light third generation of sfermions. Moreover, in the limit when both
solutions are ``perfect'' and assure negligible contributions from the
squark flavour mixing, the only potentially significant contribution
to FCNC transitions may come from the (charged Higgs)-top and
chargino-stop loops with Yukawa couplings and KM angles in the
vertices.\footnote{
Chargino-sbottom loops could be important in \dd mixing as well. We
will not discuss this possibility here, although it could be
well motivated in large $\tan\beta$ scenarios.}
Since, in addition, several arguments (see section~\ref{sec:rge})
suggest that charginos and 3rd generation of sfermions may be among
the lightest superpartners, it is interesting to discuss in more
detail their impact on FCNC transitions.

	The present section is devoted to discussing such a
scenario. In the first step, the only extra MSSM contributions to the
FCNC processes we consider are the (charged Higgs)-top and
chargino-stop loops. Our results depend then on the following
parameters (apart from the SM ones):
\begin{itemize}
\item[(i)] $\tan \beta$
\item[(ii)] Physical masses of the lighter and heavier stop
($m_{\tilde{T}_1}$ and $m_{\tilde{T}_2}$, respectively), as well as
their mixing angle $\theta_{LR}$. The sign convention for this angle
is fixed by requiring that $(Z_U)^{63} \simeq \sin \theta_{LR}$.
\item[(iii)] Chargino mass and mixing parameters. We choose the
lightest chargino mass $m_{\chi^{\pm}_1}$ and the ratio $M_2/\mu$ as
input parameters.
\item[(iv)] The charged Higgs boson mass $m_{H^{\pm}}$.
\end{itemize}

	In most of the numerical examples, we will decouple the
heavier stop and assume that the lighter one is dominantly right,
i.e. that $\theta_{LR}$ is relatively small (of order $10^o$). This is
motivated by studies of supersymmetric effects in electroweak
precision observables \cite{CP96.PL,CP96.NP}.

	In the considered MSSM scenario, various FCNC processes
exhibit different sensitivity to supersymmetry. While sizable effects
can still occur in the neutral meson mixing (\kk and \bb),
supersymmetric contributions to other FCNC processes are usually
either small or screened by long-distance QCD effects. An exception is
the inclusive weak radiative B meson decay $B \to X_s \gamma$, to
which light superpartners can contribute significantly, and where
strong interaction effects are under control. In the following, we
shall first focus on neutral meson mixing and then discuss the $B \to
X_s \gamma$ decay.

	In the considered approach to the MSSM, the results for $\Delta
m_{B_d}$ and $\epsilon_K$ read
\bea
\Delta m_{B_d} &=& \eta_{\scs QCD}  {\alpha_{em}^2 m_t^2 \over 12 
\sin^4\theta_W M_W^4} f_{B_d}^2 B_{B_d} m_{B_d}
|K_{tb}K_{td}^\star|^2| \Delta|,
\label{eq:dmbd}\\
|\epsilon_K| &=&{\sqrt{2}\alpha_{em}^2 m_c^2\over 48 
\sin^4\theta_W M_W^4} f_K^2 B_K 
{m_K\over \Delta m_K} |{\cal I}m\Omega|,
\label{eq:epsk}
\eea
where
\bea
\Omega = \eta_{cc}(K_{cs}K_{cd}^\star)^2
+ 2\eta_{ct}(K_{cs}K_{cd}^\star K_{ts}K_{td}^\star)
f\left(\frac{m_c^2}{M_W^2},\frac{m_t^2}{M_W^2}\right)
+\eta_{tt}(K_{ts}K_{td}^\star)^2\frac{m_t^2}{m_c^2}\Delta, 
\label{eq:epskom}
\eea
and
\bea
f(x,y) = \log {y\over x} 
+ {3y\over 4(y-1)}\left(1 - {y\over y -1}\log y\right).\nonumber
\eea

	The charged Higgs and the chargino boxes enter, together with
the SM terms, only into the quantity $\Delta$ in the above
equations. The QCD correction factors $\eta_{cc}$, $\eta_{ct}$,
$\eta_{tt}$ and $\eta_{\scs QCD}$ are known up to the next-to-leading
accuracy \cite{etaNLO}.

	The KM elements appearing in eqs.~(\ref{eq:dmbd}-\ref{eq:epskom})
can be conveniently expressed in terms of the Wolfenstein parameters
$\lambda$, $A$, $\rho$ and $\eta$~\cite{WOLFENSTEIN}
\bea
K\approx\left(
\begin{array}{ccc}
1-\frac{\lambda^2}{2} & \lambda & A\lambda^3(\rho-i\eta)\\
- \lambda-iA^2\lambda^5\eta& 
             1-\frac{\lambda^2}{2} + i{\cal O}(\lambda^6) & A\lambda^2\\
 A\lambda^3(1-\rho-i\eta)&-A\lambda^2 -iA\lambda^4\eta & 1\\
\end{array}
\right) + {\cal O}(\lambda^4),
\label{eq:wolf}
\eea
where $\lambda = 0.22$ is known from semileptonic kaon and hyperon
decays. The leading in $\lambda$ imaginary parts of all the entries
are shown, because most of them are relevant in analyzing CP-violation.

	Both in the SM and MSSM, the theoretical predictions for \epsk
and \dmbd have some uncertainty due to non-perturbative parameters
$B_K$, $f_{B_d}^2 B_{B_d}$ which are known from lattice calculations,
but not very precisely. Moreover, the KM element $K_{td} =
A\lambda^3(1-\rho-i\eta)$ which appears in
eqs.~(\ref{eq:dmbd}-\ref{eq:epskom}) is not directly measured. Its SM
value fitted to the observables in eqs.~(\ref{eq:dmbd}-\ref{eq:epsk})
can change after inclusion of new contributions. Thus, the correct
approach is to fit the parameters $A$, $\rho$, $\eta$ and $\Delta$ in
a model independent way to the experimental values of \epsk and \dmbd
\cite{BFZ96}. The quantities $|K_{cb}|$ and $|K_{ub}/K_{cb}|$ are
known from tree level processes. They are practically unaffected by
new physics which contributes only at one and more loops.

	Here, we give the results of such a fit, with $B_K$ and
$f_{B_d}^2 B_{B_d}$ varied in a the following ranges:~\cite{B96}.
\bea
0.6<&B_K&<0.9
\label{eq:bkrange}
\\
0.160~\mathrm{GeV}<&\sqrt{f_{B_d}^2 B_{B_d}}&<0.240~\mathrm{GeV}
\label{eq:fbrange}
\eea
In our fit, we use the following experimental results~\cite{B96,PDG96}:
\bea 
|K_{cb}| &=& 0.039\pm 0.002
\label{eq:expkcb}\\
|K_{ub}/K_{cb}|&=&0.08\pm 0.02
\label{eq:expkrat}\\
|\epsilon_K| &=&(2.26\pm 0.02)\, 10^{-3}
\label{eq:expepsk}\\
\dmbd &=& (3.01\pm 0.13)\, 10^{-13}~\mathrm{GeV}
\label{eq:expdmb} 
\eea

\begin{figure}[htbp]
\begin{center}
\begin{tabular}{p{0.48\linewidth}p{0.48\linewidth}}
\mbox{\epsfig{file=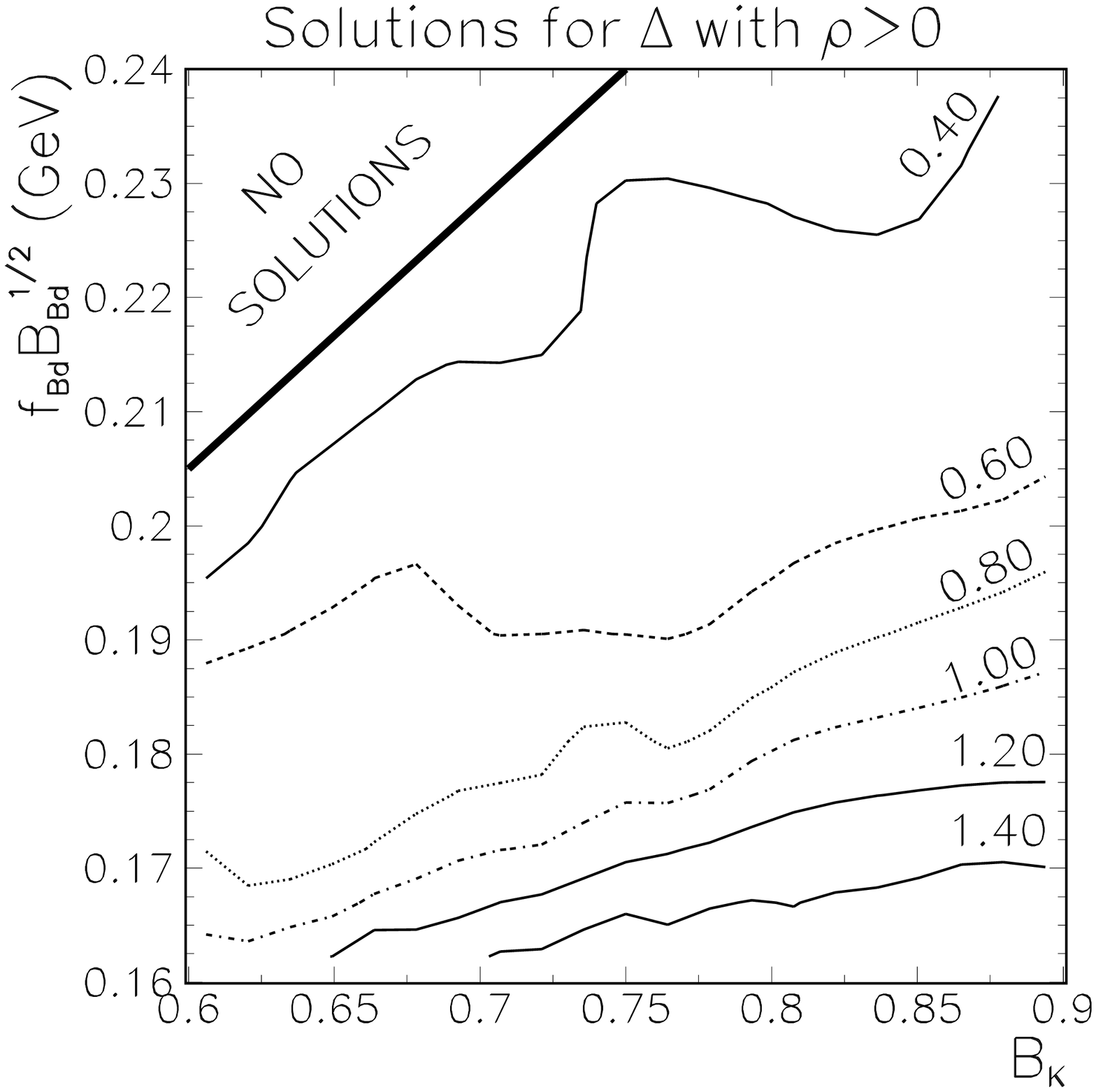,width=\linewidth}}
&
\mbox{\epsfig{file=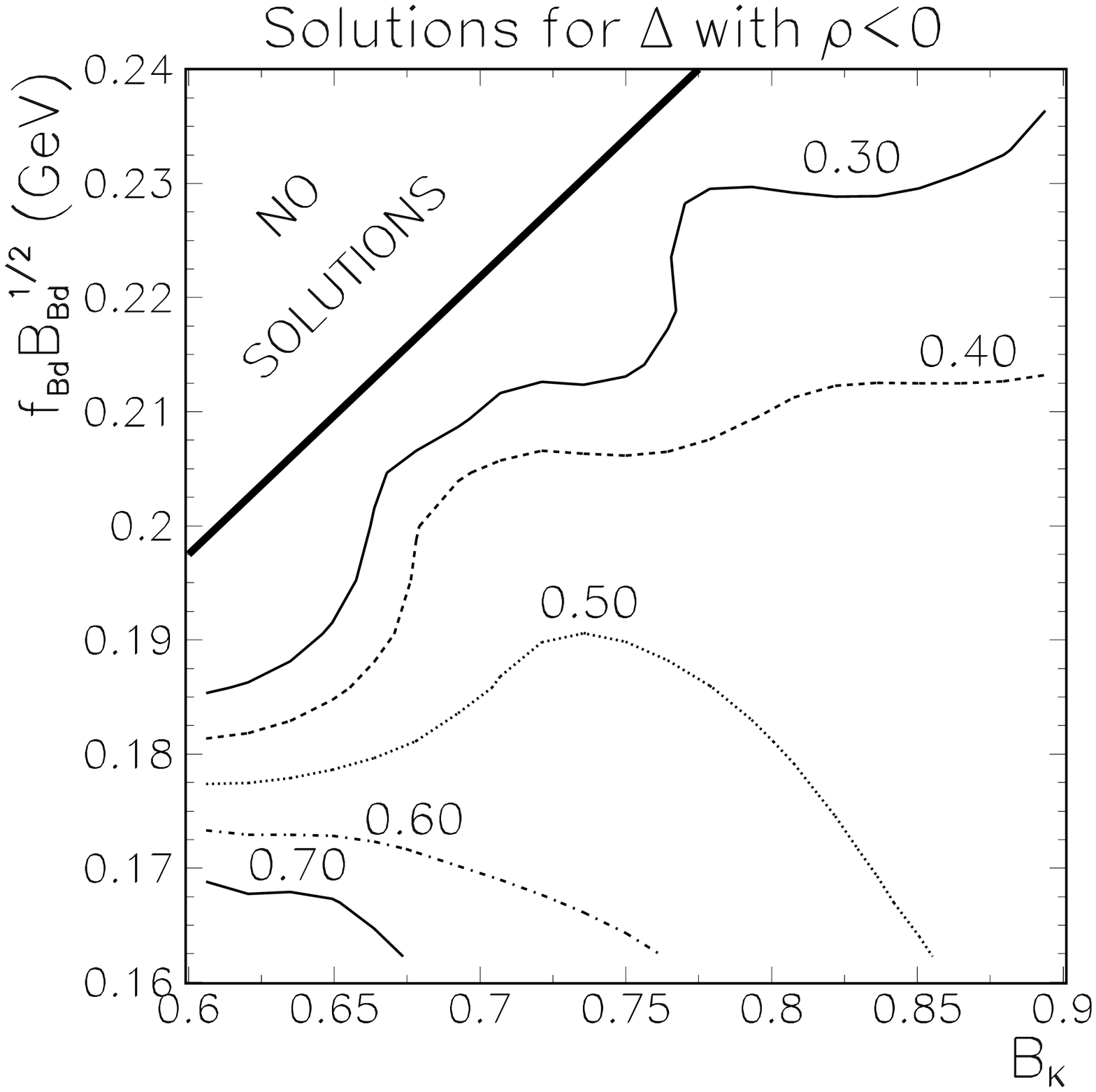,width=\linewidth}}
\\
\end{tabular}
\caption{Contour lines of the parameter $\Delta$ minimizing the 
$\chi^2$ fit to experimentally measured values of $|K_{cb}|$,
$|K_{ub}/K_{cb}|$, \epsk and \dmbd.\label{fig:delfit}}
\end{center}
\end{figure}

	In Fig.~\ref{fig:delfit}, we show values\footnote{
Here, we assume that $\Delta$ is real. This is true in the SM and, to
a very good approximation, in the considered approach to the
MSSM. However, in a general MSSM, $\Delta$ could develop a sizable
imaginary part.}
of the parameter $\Delta$ obtained from the $\chi^2$ fit of the
parameters $(A,\rho,\eta,\Delta)$ to the four quantities listed in
eqs.~(\ref{eq:expkcb}-\ref{eq:expdmb}), as a function of $B_K$ and
$f_{B_d} (B_{B_d})^{1/2}$. The left and right plots in
Fig.~\ref{fig:delfit} are equivalent to two different solutions for
$\chi^2$ minimum with the Wolfenstein parameter $\rho>0$ and $\rho<0$,
respectively. In our fit, we require $\chi^2_{min}\leq 4$. As can be
seen from both plots of Fig.~\ref{fig:delfit}, no such solutions exist
for small $B_K$ and large $f_{B_d} (B_{B_d})^{1/2}$, where
$\chi^2_{min}$ starts to grow quickly. In the remaining $(B_K,f_{B_d}
(B_{B_d})^{1/2})$ range, $\chi^2_{min}$ is close or equal to 0,
excluding only those values of $(B_K,f_{B_d} (B_{B_d})^{1/2})$ which
are very close to the thick boundary line marked in both plots. The
contour lines show values of $\Delta$ which exactly minimize the
$\chi^2$ fit. The $1\sigma$ errors on $\Delta$ obtained from the fit
are typically of the order of ${\cal O}(0.1-0.2)$, depending on the
specific values of $B_K$ and $f_{B_d} (B_{B_d})^{1/2}$. Results
plotted in Fig.~\ref{fig:delfit} can be compared with the theoretical
prediction for the parameter $\Delta$ in the SM: $\Delta_{SM} \approx
0.53$. As can be seen from Fig.~\ref{fig:delfit}, larger values of
$\Delta>\Delta_{SM}$ (interesting in the MSSM, as discussed later)
prefer $\rho>0$, small values of $f_{B_d} (B_{B_d})^{1/2}$ and, to a
lesser extent, large $B_K$. For instance, $\Delta>1$ requires $\rho>0$
and $f_{B_d} (B_{B_d})^{1/2}<0.19~\GeV$. Scanning over allowed range
for $B_K$ and $f_{B_d} (B_{B_d})^{1/2}$, defined in
eqs.~(\ref{eq:bkrange}-\ref{eq:fbrange}), gives the ``absolute''
bounds on $\Delta$. Such bounds are not very tight. After including
$1\sigma$ errors on $\Delta$, they are roughly
\bea
0.2 \simleq \Delta \simleq 2.0
\label{eq:dlimit}
\eea

\begin{figure}[htbp]
\begin{center}
\mbox{\epsfig{file=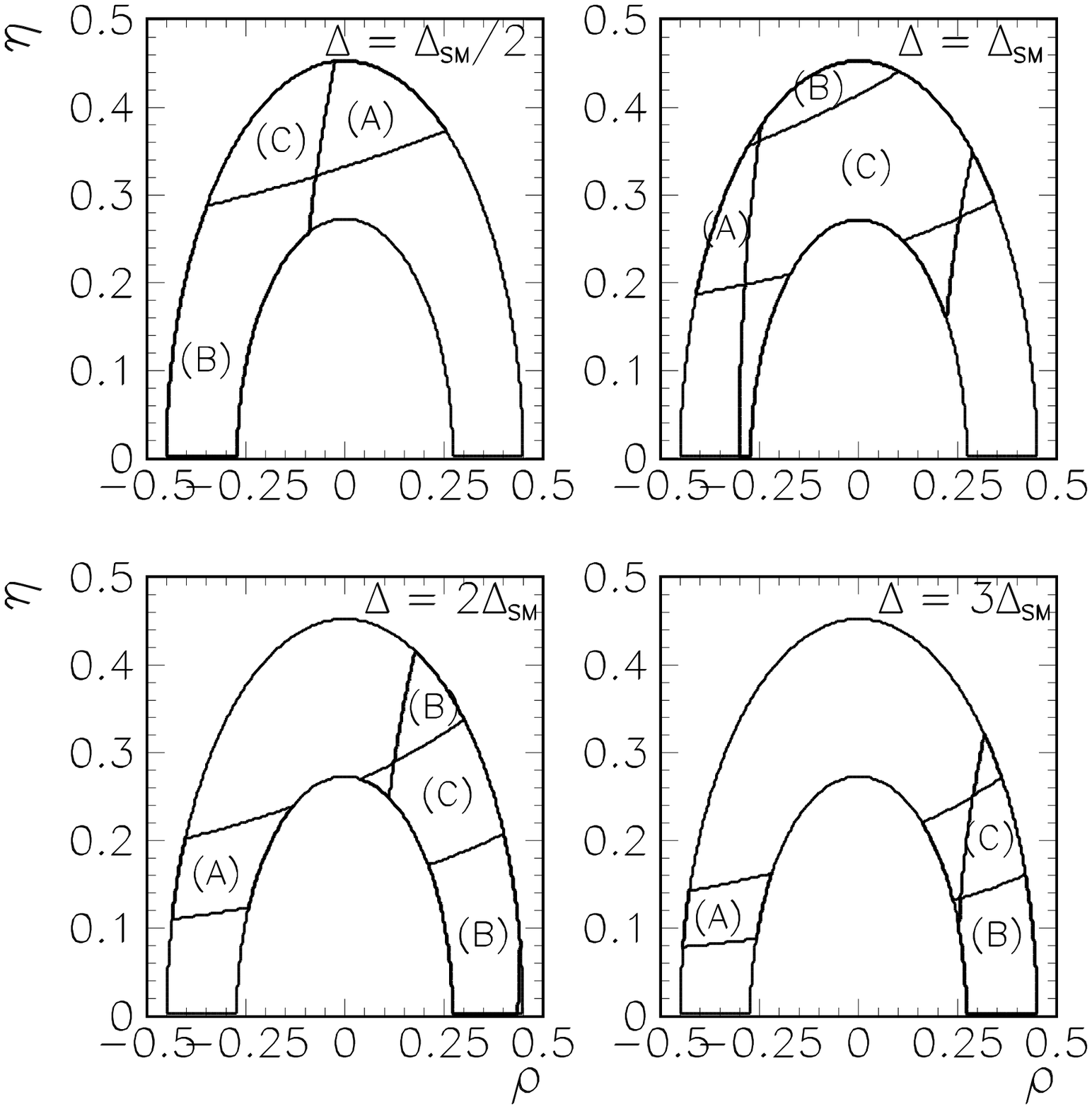,width=\linewidth}}
\caption{Allowed regions in the $(\rho,\eta)$ plane for four values of
$\Delta$: (A) - allowed by $\epsilon_K$, (B) - allowed by $\Delta
m_{B_d}$, (C) - allowed by \epsk and $\Delta
m_{B_d}$. The value of $\Delta_{SM}$ is
approximately equal to 0.53.\label{fig:erfit}}
\end{center}
\end{figure}

	In Fig.~\ref{fig:erfit}, we plot the allowed ranges of $\rho$ and
$\eta$ for several fixed values of
$\Delta=\frac{1}{2}\Delta_{SM},\Delta_{SM},2\Delta_{SM},3\Delta_{SM}$
and changing $B_K$, $f_{B_d} (B_{B_d})^{1/2}$ in the ranges specified
in eqs.~(\ref{eq:bkrange}-\ref{eq:fbrange}). The allowed half-ring
visible in the plots of Fig.~\ref{fig:erfit} originates from
$|K_{ub}/K_{cb}|$ given in eq.~(\ref{eq:expkrat}). The measurement of
\dmbd allows another ring in the $(\rho,\eta)$ plane. It is centered
at $(\rho,\eta) = (1,0)$. Its interesting part is approximately
parallel to the $\eta$ axis. It moves to the right (towards larger
$\rho$) when $\Delta$ increases. The range bounded by \epsk is
approximately parallel to the $\rho$ axis. It moves down (towards
smaller $\eta$) with increasing $\Delta$. Taking both effects into
account, we can see that small $\Delta$ prefers negative $\rho$ and
large $\eta$, $\Delta\sim\Delta_{SM}$ gives the biggest allowed range
for $\rho$ and $\eta$ with both $\rho<0$ and $\rho>0$ possible,
whereas larger $\Delta\geq 1$ requires positive $\rho$ and
smaller~$\eta$.

	In the next step, we correlate the value of $\Delta$ with 
\begin{figure}[htbp]
\begin{center}
\mbox{\epsfig{file=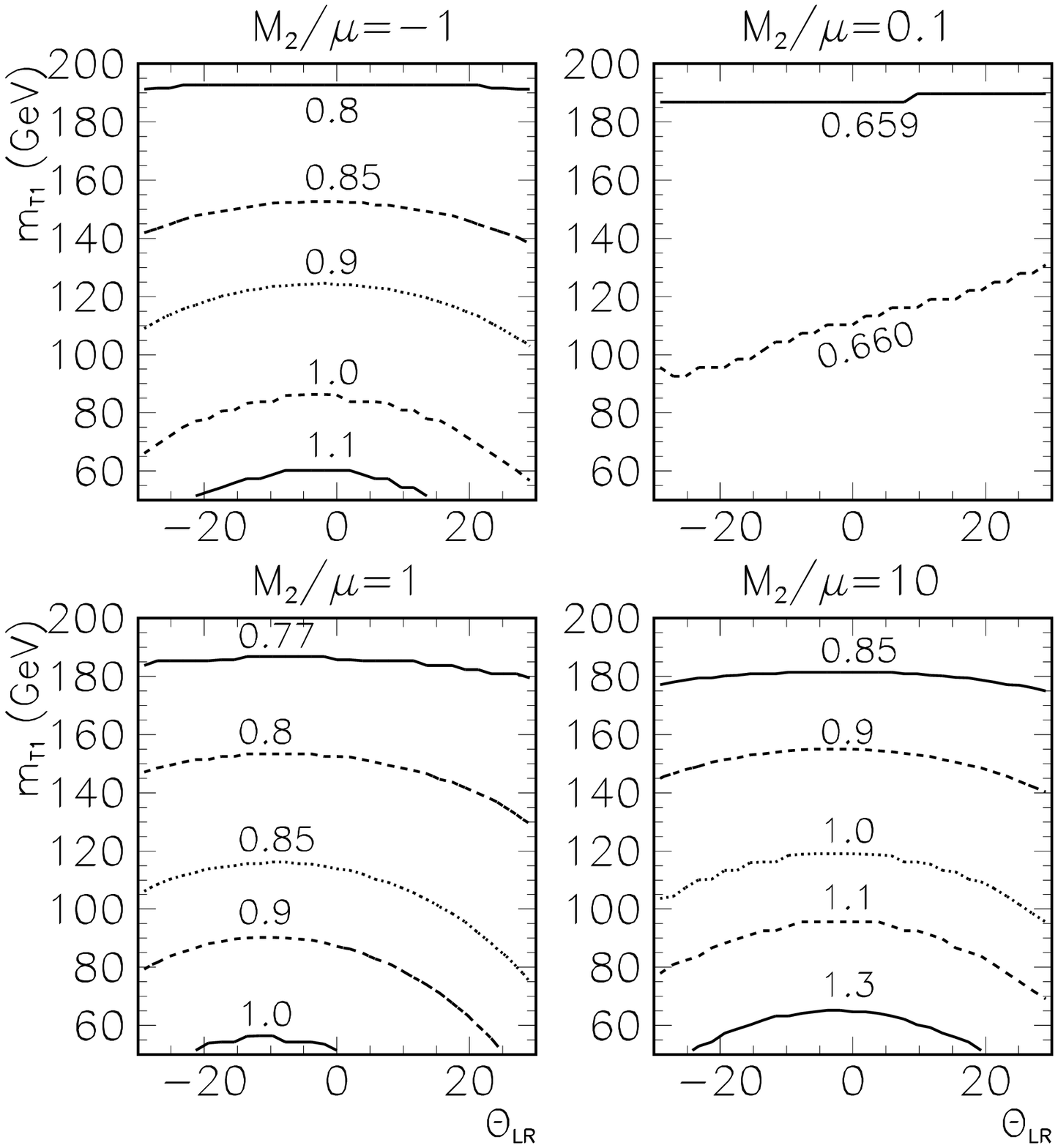,width=\linewidth}}
\end{center}
\caption{Contour lines of $\Delta$ as a function of right stop mass
and stop mixing angle for $\tan\beta=1.8$, $M_{H^+}= 100$ GeV,
$M_{\tilde{T}_2}=250$ GeV, $m_{\chi^-}=90$ GeV and four chosen
$M_2/\mu$ ratio values.}
\label{fig:rsfit}
\end{figure}
masses and mixings in the MSSM. In Fig.~\ref{fig:rsfit}, we plot
contour lines of constant $\Delta$ for light SUSY spectrum, i.e. in
the range where SUSY effects are most visible. As seen from
Fig.~\ref{fig:rsfit}, the values of $\Delta$ in the MSSM are always
bigger than in the SM, i.e. the new contributions to $\Delta$ from the
Higgs and chargino sectors have the same sign as $\Delta_{SM}\approx
0.53$. This is a general conclusion, always true for the Higgs
contribution and valid also for the chargino-stop contribution when
SUSY parameters are chosen as in this section.  The actual value of
the supersymmetric contribution to $\Delta$ depends strongly on the
ratio $M_2/\mu$. The charged Higgs contribution does not depend on
this ratio, and increases $\Delta$ by about 0.12 for $m_{H^\pm} = 100$
GeV and $\tan\beta=1.8$, as used in Fig.~\ref{fig:rsfit}. For small
values of $|M_2/\mu|$, when the lighter chargino is predominantly
gaugino, the $\chi^-$--$\tilde{T}_1$ contribution to $\Delta$ is very
small (of order $10^{-2}$) and weakly dependent on the lighter stop
mass. This can be easily understood: In this case, the lighter stop is
coupled to the lighter chargino mostly through the LR mixing in the
stop sector, and the appropriate contribution is suppressed by
$\sin^4\theta_{LR}$. For larger values of $|M_2/\mu|\sim 1$, this
contribution is bigger and, due to the interference between the
diagrams with and without the LR mixing, may reach its maximal value
for $\theta_{LR}\neq 0$, depending on the sign of
$\mu$. Chargino-lighter stop contributions increase further with
$|M_2/\mu|$, when lighter chargino consists predominantly of Higgsino,
and become again independent on the sign of $\mu$. In this case,
contributions proportional to $Z_U^{63}\approx \sin\theta_{LR}$ are
negligible, and varying of $\theta_{LR}$ is visible in $\Delta$ only
via $Z_U^{66} \simeq \cos\theta_{LR}$.

	Increasing the charged Higgs mass to $m_{H^{\pm}}\approx 500$
GeV and chargino mass to $m_{\chi_1^{\pm}}=300$ GeV suppresses the
magnitude of each contribution by a factor of 3 approximately, but
does not change the character of its dependence on $\theta_{LR}$. The
results illustrated in Fig.~\ref{fig:rsfit} are also weakly dependent
on the mass of the left stop: Increasing $M_{\tilde{T}_2}$ from 250 to
500 GeV modifies $\Delta$ only marginally.

	We now turn to the discussion of $B \to X_s \gamma$.
Sample SM and MSSM diagrams for the $b \to s \gamma$ transition
are shown in Figs.~\ref{fig:SMbsg} and~\ref{fig:SUSYbsg}, respectively

\input bsg_diag.axo

Dressing these diagrams with one or more gluons gives us QCD
contributions enhanced by large logarithms $\ln(M_W^2/m_b^2)$. In the
SM, they increase the decay rate by more than a factor of 2. Resumming
these large QCD logarithms up to next-to-leading order (NLO) is
necessary to acquire sufficient accuracy \cite{BMMP94}. Such a
resummation has been recently accomplished in the SM
\cite{CMM96,GHW96,AY94,AG95}.

	The analysis of $B \to X_s \gamma$ decay begins with
introducing an effective Hamiltonian
\be \label{heff}
H_{eff} = 
-\f{4 G_F}{\sqrt{2}} K^*_{ts} K_{tb} \sum_{i=1}^{8} C_i(\mu) P_i(\mu) 
\ee
where $P_i$ are the relevant operators and $C_i(\mu)$ are their
Wilson coefficients. Here, we need to give only two of these
operators explicitly
\bea 
P_7  &=&  \f{e}{16 \pi^2} m_b (\bar{s}_L \sigma^{\mu \nu}     b_R) F_{\mu \nu} 
\label{P7} \\  
P_8  &=&  \f{g_3}{16 \pi^2} m_b (\bar{s}_L \sigma^{\mu \nu} T^a b_R) 
G_{\mu \nu}^a \label{P8}
\eea
where $F_{\mu\nu}$ and $G_{\mu\nu}^a$ are the photonic and gluonic
field strength tensors, respectively.  Resummation of large logarithms
$\ln(M_W^2/m_b^2)$ is achieved by evolving the coefficients $C_i(\mu)$
from $\mu \sim M_W$ to $\mu \sim m_b$ according to the renormalization
group equations.  Feynman rules derived from the effective Hamiltonian
are then used to calculate the $b$-quark decay rate $\Gamma[b \to X_s
\gamma]$ which is a good approximation to the corresponding $B$-meson
decay rate \cite{FLS94,N94}. All these calculations are identical in
the SM and MSSM (also at NLO), except for that the initial numerical
values of the Wilson coefficients $C_7$ and $C_8$ at $\mu \sim M_W$
are different\footnote{\noindent
It would not be the case for arbitrary MSSM parameters when
extra operators in the effective theory could arise. However, so
long as our assumptions from the beginning of this section are
fulfilled, all these extra operators are negligible.}.
The leading-order MSSM contributions to $C_7(M_W)$ and $C_8(M_W)$ are
well known \cite{BBMR91,CMW96}. Here, we quote only the SM, charged
higgson and chargino contributions to $C_7(M_W)$
\bea 
C_7^{(0)SM}(M_W) &=& \f{1}{4} \f{m_t^2}{M_W^2} f_1\left( \f{m_t^2}
{M_W^2} 
\right) 
\label{eq:c7SM}\\
C_7^{(0)H^{\pm}}(M_W) &=& 
\f{1}{12} \f{m_t^2}{M_{H^{\pm}}^2} \cot^2 \beta 
f_1\left(\f{m_t^2}{M_{H^{\pm}}^2}\right)
+ \f{1}{6} f_2\left( \f{m_t^2}{M_{H^{\pm}}^2} \right)
\label{eq:c7Hpm}\\
C_7^{(0)\chi^{\pm}}(M_W) &=& 
\f{1}{K_{ts}^* K_{tb}} \sum_{i=1}^6 \sum_{p=1}^2 \f{M_W^2}
{m_{\chi_p^{\pm}}^2} 
\left\{ 
-\f{1}{6} A^{p*}_{i2} A^p_{i3} 
f_1\left( \f{m_{\tilde{U}_i}^2}{m_{\chi_p^{\pm}}^2}\right)
\right.\nonumber\\ 
&+&\left.\f{1}{3}A^{p*}_{i2} B^p_{i3} \f{m_{\chi_p^{\pm}}}{m_b}
f_2\left( \f{m_{\tilde{U}_i}^2}{m_{\chi_p^{\pm}}^2} \right) 
\right\}
\label{eq:c7ch}
\eea
where 
\bea
f_1(x) &=& \f{ 3 x^2 - 2 x}{(1-x)^4} \log x + \f{ 8 x^2 + 5 x - 7}
{6(1-x)^3},
\nonumber\\
f_2(x) &=& \f{-3 x^2 + 2 x}{(1-x)^3} \log x + \f{-5 x^2 + 3 x    }
{2(1-x)^2}.
\nonumber
\eea
The matrices $A$ and $B$ originate from the vertices in 
Fig.~\ref{fig:ccurrent}
\bea
A^p_{kJ} &=& \left[ - Z_U^{Ik*} Z^+_{1p} 
   + \f{1}{\sqrt{2} M_W \sin \beta} m_u^{II} Z_U^{(I+3)k*} Z^+_{2p} 
\right] K^{IJ},\nonumber \\
B^p_{kJ} &=& \f{1}{\sqrt{2} M_W \cos \beta} Z_U^{Ik*} K^{IJ} m_d^{JJ} 
Z^{-*}_{2p}.
\nonumber
\eea

	The next-to-leading corrections to $C_7(M_W)$ are found by
taking diagrams from Figs.~\ref{fig:SMbsg} and~\ref{fig:SUSYbsg},
adding one virtual gluon to them, and calculating their short-distance
part. This has been done only in the SM case~\cite{AY94}. In the
supersymmetric case, only contributions proportional to logarithms of
superpartner masses are known~\cite{A94}, which is enough only for
very heavy superpartners.

	So long as the superpartners are heavy and their contributions
to $C_7(M_W)$ are small (say, below 30\%), then it does not really
matter that the SUSY NLO corrections to $C_7(M_W)$ are unknown. The
uncertainty is then dominated by SM sources anyway.  However, if some
of the SUSY contributions are big but cancel each other, so that the
experimental $B \to X_s \gamma$ constraints \cite{CLEO95} are
fulfilled, then the lack of SUSY NLO corrections to $C_7(M_W)$ does
matter. Such cancellations really do occur in sizable and interesting
domains of the MSSM parameter space \cite{BG93,CP96.NP}.  This is why a
calculation of the SUSY NLO corrections to $C_7(M_W)$ would be
welcome.

	In the results presented below, the complete Standard Model
NLO formulae and values of parameters are used precisely as they stand
in ref.~\cite{CMM96}. As far as SUSY contributions to $C_7(M_W)$ and
$C_8(M_W)$ are concerned, we have used only the available leading
order results.  The charged higgson and chargino contributions to
$C_7(M_W)$ are separately assumed to have additional 10\% uncertainty
due to order $\alpha_s(M_W)/\pi$ corrections to them. These
uncertainties are added in squares to the remaining errors. This way
we have simulated growth of uncertainty in cases where large
cancellations between SUSY contributions occur.
\begin{figure}[htbp]
\begin{tabular}{p{0.48\linewidth}p{0.48\linewidth}}             
\begin{center}
\mbox{\epsfig{file=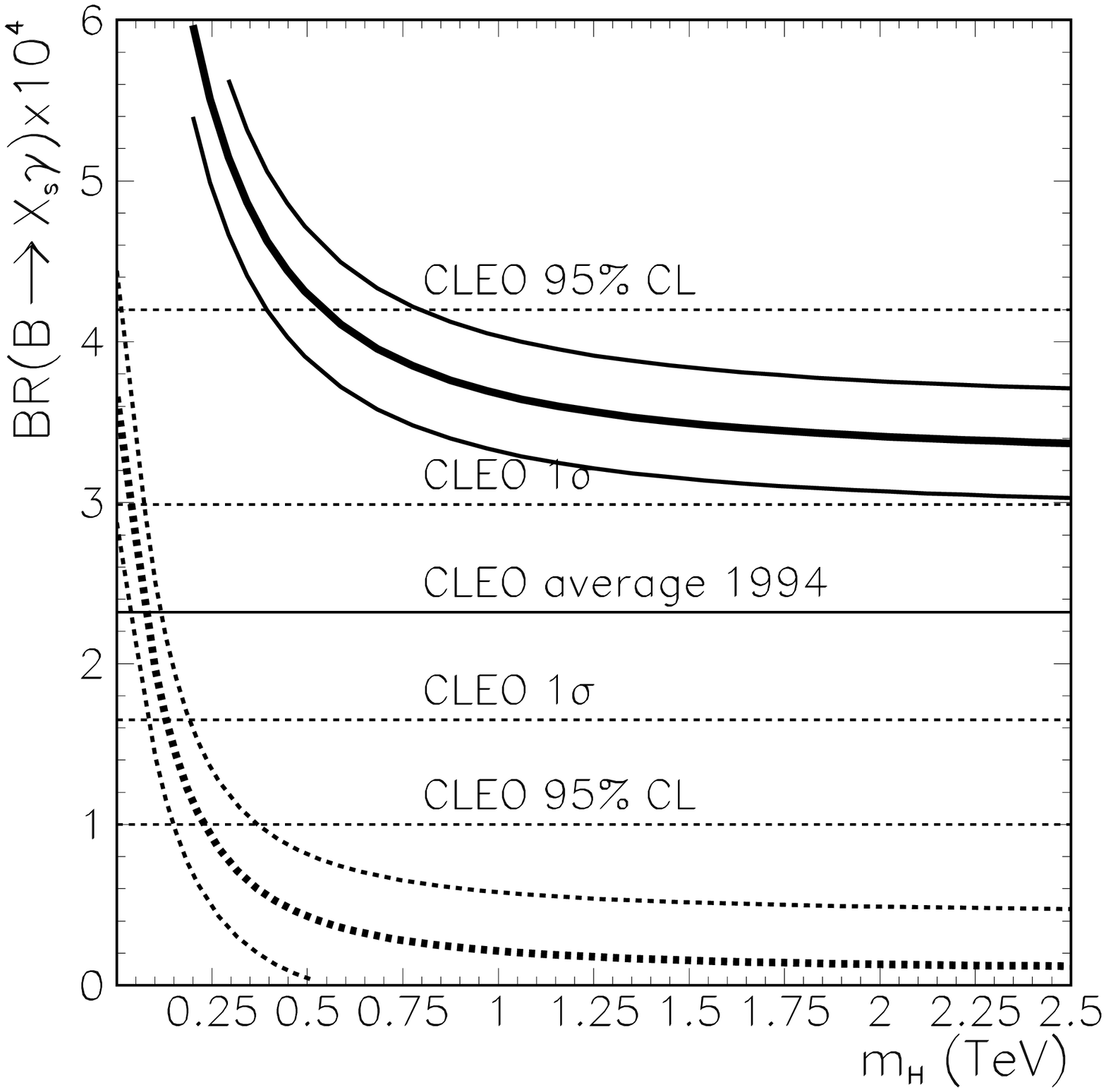,width=\linewidth}}
\caption{$Br[B \to X_s \gamma]$ as a function of the charged 
higgson mass.}
\label{fig:bsg_mh}
\end{center}
&
\begin{center}
\mbox{\epsfig{file=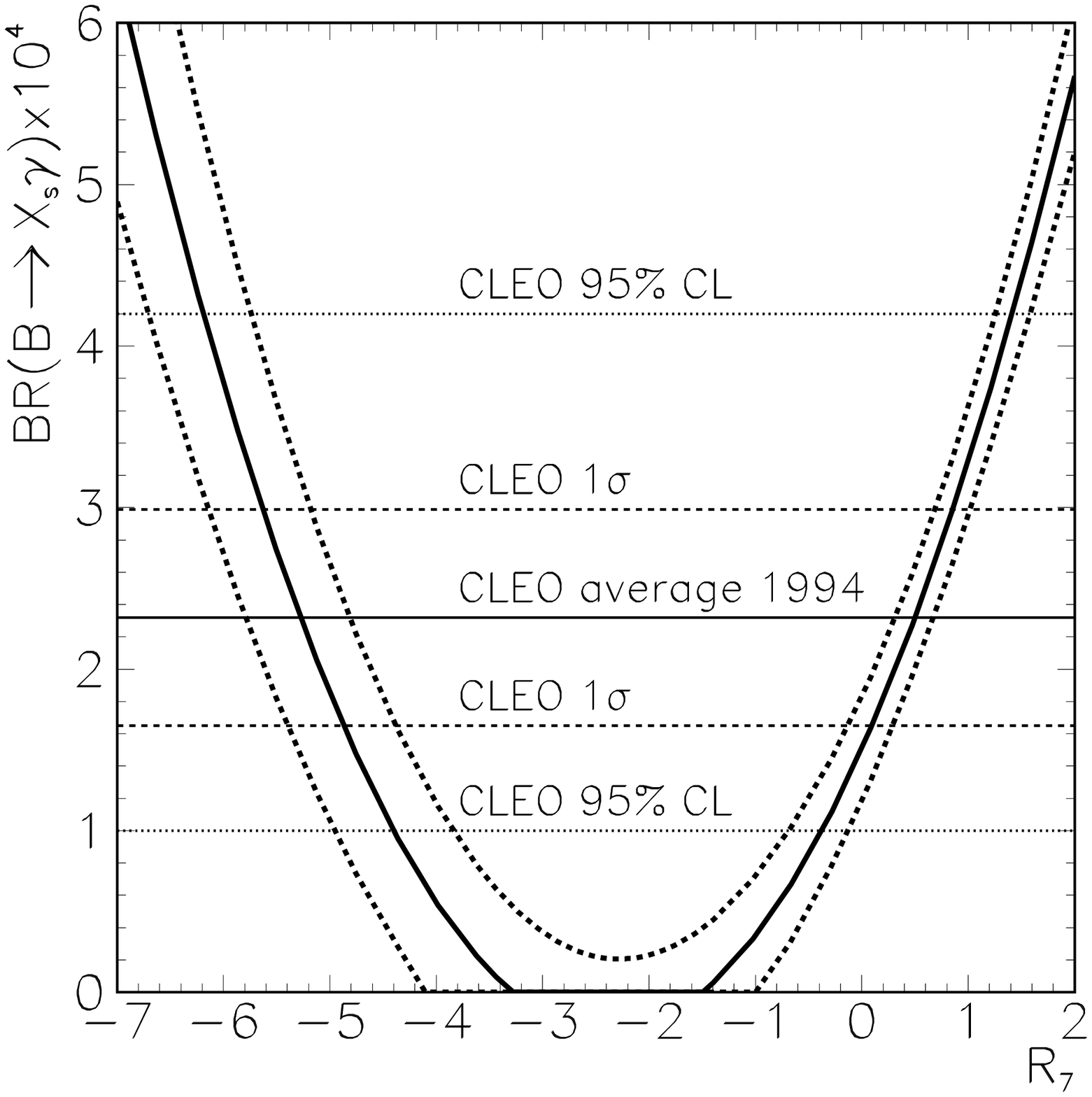,width=\linewidth}}
\caption{$Br[B \to X_s \gamma]$ as a function of $R_7$.}
\label{fig:r7}
\end{center}
\end{tabular}             
\end{figure}

	Figure~\ref{fig:bsg_mh} presents two examples of how $Br[B \to
X_s \gamma]$ depends on the charged higgson mass. Solid lines
correspond to the case when all the superpartner masses are large
(above 1~TeV). In this case, the MSSM results are the same as in the
Two-Higgs-Doublet Model II. The value of $\tan \beta = 2$ was used.
However, any larger value would give almost the same results, while
for smaller $\tan \beta$, we would have an additional enhancement. The
middle line corresponds to the central value, and the two remaining
lines show the estimated uncertainty. For heavy charged higgson (well
above 1 TeV), the curve approaches the SM result $Br[B \to X_s \gamma]
= (3.28 \pm 0.33) \times 10^{-4}$. The horizontal lines show the CLEO
$1\sigma$ error bar $(2.32 \pm 0.67) \times 10^{-4}$ as well as their
95\% CL upper and lower bounds \cite{CLEO95}. We can see that the
theoretical prediction crosses the 95\% CL upper line close to
$M_{H^{\pm}} = 500$ GeV. This sets the lower bound on this mass in the
considered MSSM scenario, and an absolute lower bound in the
Two-Higgs-Doublet Model II.

	Dashed lines in Fig.~\ref{fig:bsg_mh} correspond to another
example. Here, chargino contributions are not negligible. We have
taken $M_{\chi^{\pm}_1} = 90$~GeV, $m_{\tilde{T}_1} = 100$~GeV, $\tan
\beta = 2$, $M_2/\mu = -5$ and $\theta_{LR} = 25^{\circ}$. All the
squarks and sleptons except $\tilde{T}_1$ are assumed to be heavier
than 1~TeV, which makes their contributions negligible. One can see
that no lower bound on the charged higgson mass can be derived from $B
\to X_s \gamma$ in this case. Actually, $H^{\pm}$ has to be relatively
light here in order to cancel the chargino contribution and bring the
prediction back to the experimentally allowed range.

	In the MSSM as well as in many other extensions of the
Standard Model (in which the NLO corrections to $C_i(M_W)$ are
unknown), extra contributions to $B \to X_s \gamma$ can be
parameterized in terms of two parameters
\be
R_7 = 1 + \f{C_7^{(0)extra}(M_W)}{C_7^{(0)SM}(M_W)},
\hspace{3cm}
R_8 = 1 + \f{C_8^{(0)extra}(M_W)}{C_8^{(0)SM}(M_W)}.
\ee
Figure~\ref{fig:r7} presents the dependence of $Br[B \to X_s
\gamma]$ on the parameter $R_7$ in the case when $R_8$ is set to
unity. The meaning of various curves is the same as in
Fig.~\ref{fig:bsg_mh}: the middle one is the central value while the
remaining two show the uncertainty. The horizontal lines are the
experimental constraints. One can see that two ranges of $R_7$
are experimentally allowed. They correspond to two possible signs
of the decay amplitude: the same or opposite than in the SM.

	The allowed ranges for $R_7$ are rather insensitive to $R_8$,
because $C_8(M_W)$ has little influence on the decay rate. For
instance, shifting $R_8$ from 1 to 3 would affect the curves in
Fig.~\ref{fig:r7} by less than the shown uncertainties. Thus, the
presented plot allows to qualitatively test various extensions of the
SM without the necessity of calculating the branching ratio itself --
it is enough to calculate $C_7^{(0)}(M_W)$ only. The situation becomes
more complex when contributions to $R_8$ are very large, as it may
happen in some corners of the MSSM parameter space \cite{CGG96}.

\begin{figure}[htbp]
\begin{center}
\begin{tabular}{p{0.48\linewidth}p{0.48\linewidth}}
\mbox{\epsfig{file=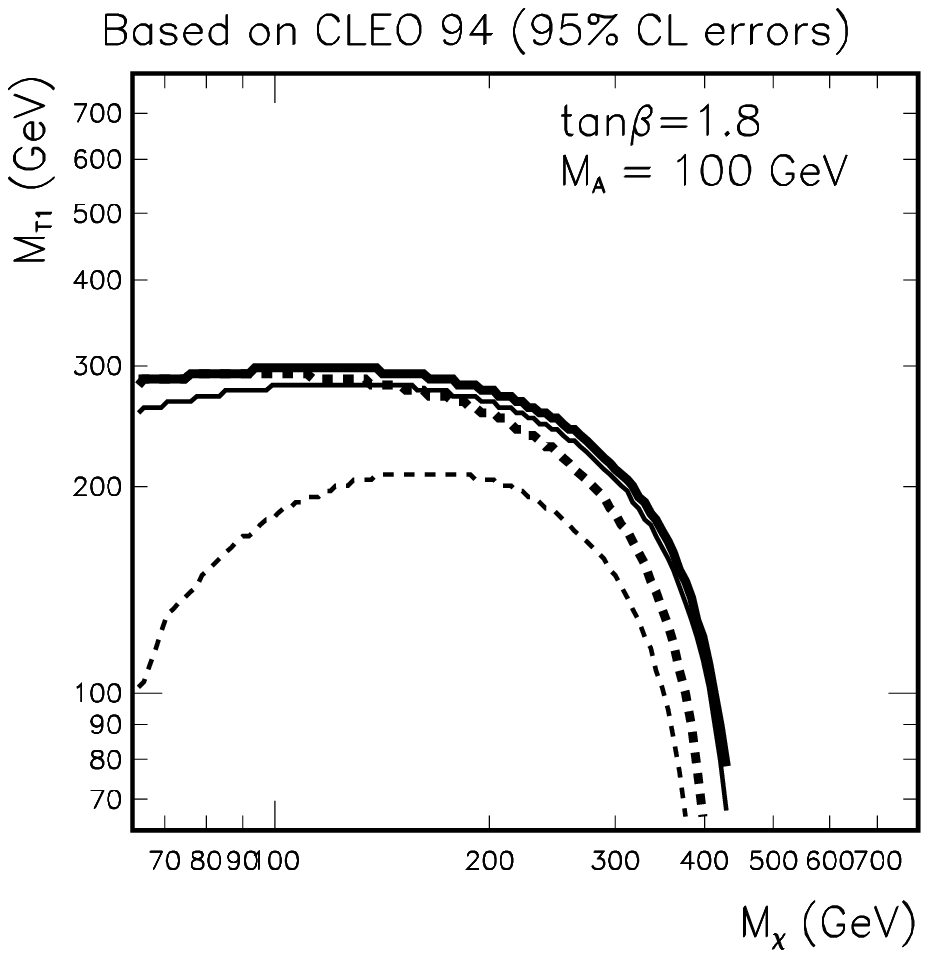,width=\linewidth}}
&
\mbox{\epsfig{file=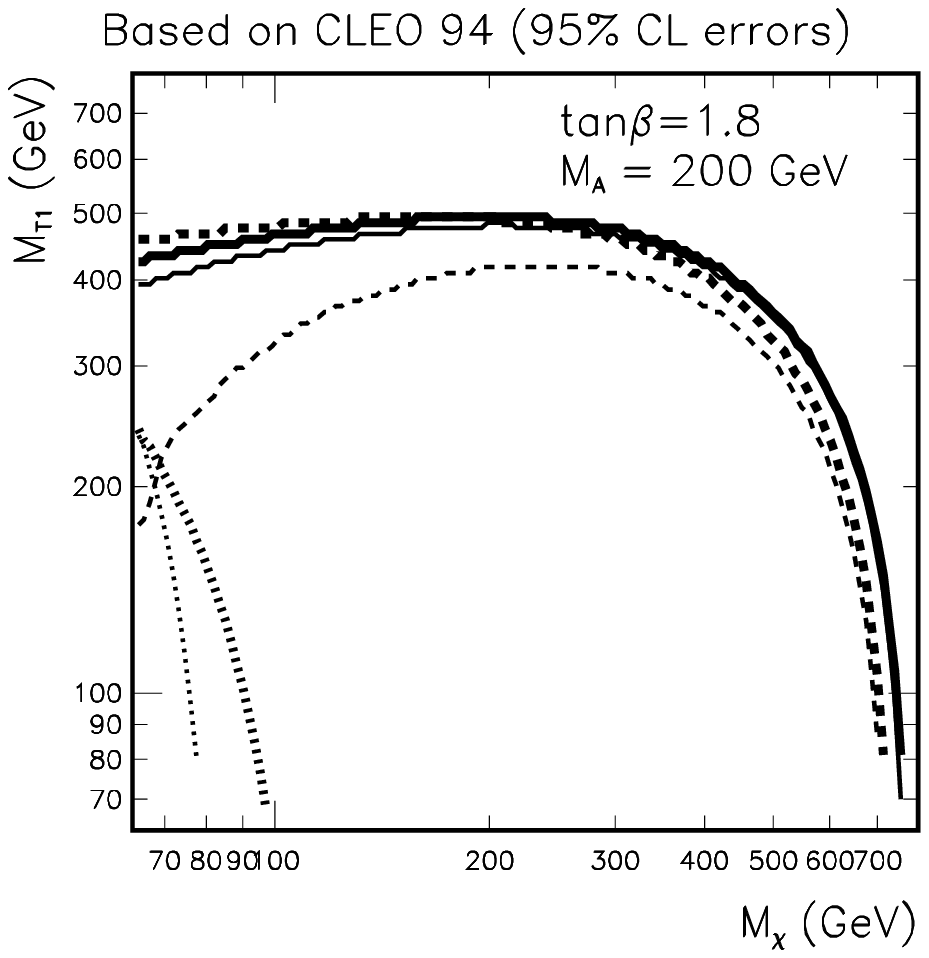,width=\linewidth}}
\\
\end{tabular}
\caption{Upper bounds on $(m_{\tilde{T_1}}$ and $m_{\chi_1^{\pm}})$
for $\tan\beta=1.8$ and $M_A=100$ and 200 GeV. Thick lines show limits
for $\mu>0$, thin lines for $\mu<0$. Dotted, dashed and solid lines
show limits for $M_2/|\mu|=0.1$, 1 and 10,
respectively.\label{fig:bsg_ct}}
\end{center}
\end{figure}

	The existing measurement of \BSG imposes already significant
constraints on the MSSM parameter space. In order to understand these
\begin{figure}[htbp]
\begin{center}
\begin{tabular}{p{0.48\linewidth}p{0.48\linewidth}}
\mbox{\epsfig{file=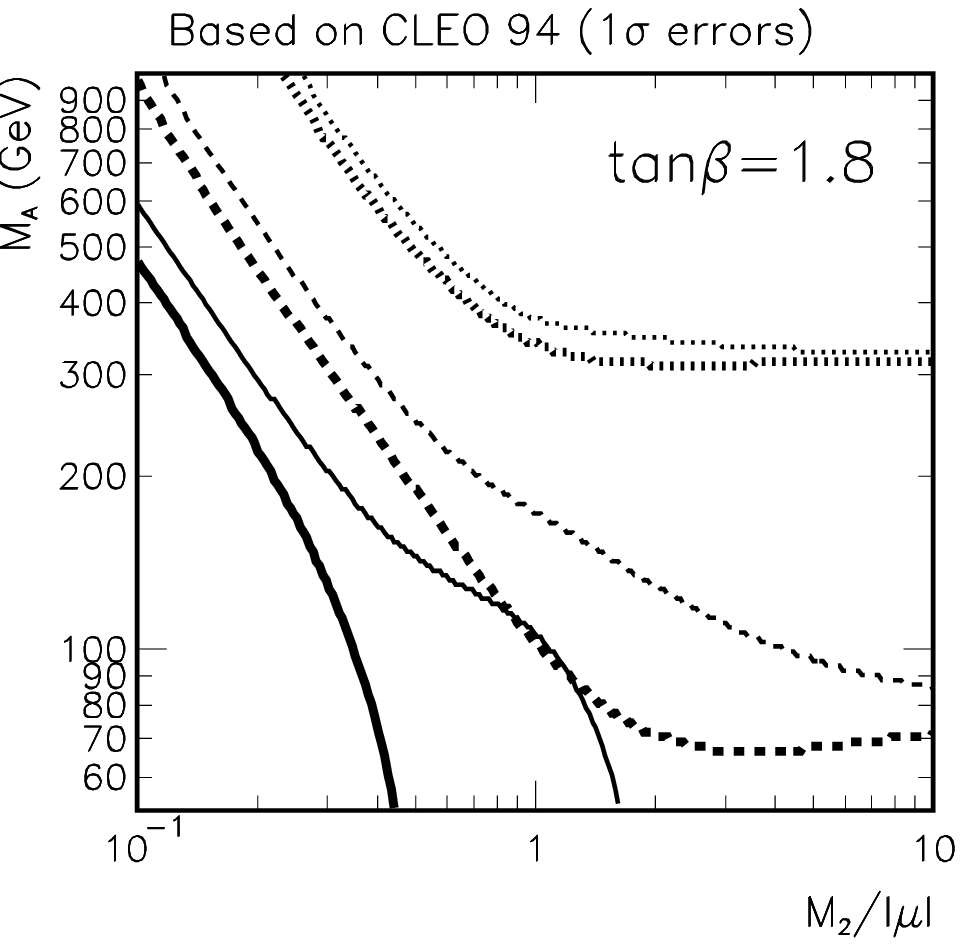,width=\linewidth}}
&
\mbox{\epsfig{file=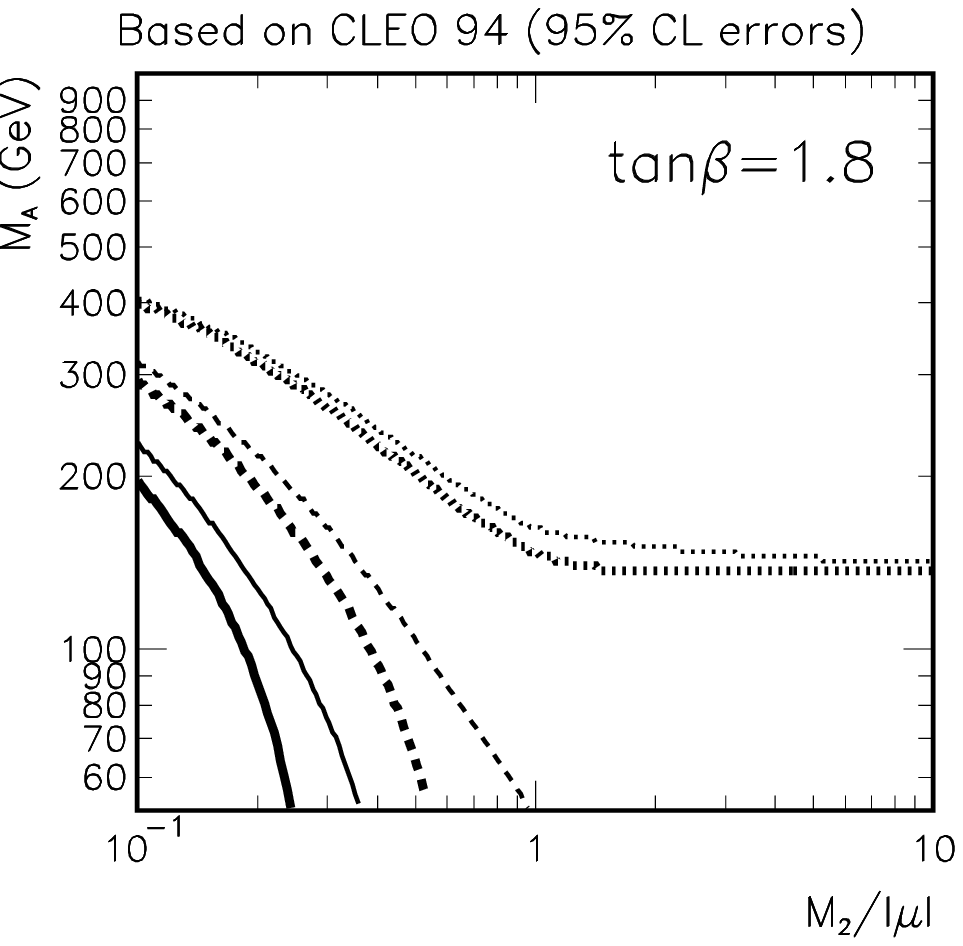,width=\linewidth}}
\\
\end{tabular}
\caption{Lower limits on allowed $M_A$ as a function of $M_2/|\mu|$,
based on CLEO $Br(B\rightarrow X_s\gamma)$ measurement. Thick lines
show limits for $\mu>0$, thin lines for $\mu<0$. Solid, dashed and
dotted lines show limits for lighter stop and chargino masses
$m_{\tilde{t}_1}=m_{\chi_1^{\pm}}=90$, 150 and 300 GeV,
respectively.\label{fig:bsg_ma}}
\end{center}
\end{figure}
limits, it is important to remember that the charged Higgs and
chargino-stop contributions to this process may have opposite
signs. Since the actually measured value of \BSG is close to the SM
prediction, SUSY contributions must either be small or cancel each
other to a large extent. Furthermore, large contributions to \BSG are
given by the Higgsino loops rather then gaugino exchanges. The content
of lighter chargino is determined by the $M_2/\mu$ parameter.  In
addition, the size of the chargino-stop contribution can be modified
by changing the stop mixing angle $\theta_{LR}$.  We illustrate those
effects in Figs.~\ref{fig:bsg_ma} and~\ref{fig:bsg_ct}.
Fig.~\ref{fig:bsg_ma} shows the lower limit on the allowed pseudoscalar
Higgs boson mass $M_A$ as a function of $M_2/|\mu|$ for three chosen
values of lighter chargino and lighter stop masses.  The charged Higgs
boson mass is in one-to-one correspondence with $M_A$: At the tree
level $M_{H^{\pm}}^2 = M_A^2 + M_W^2$. In Fig.~\ref{fig:bsg_ct}, we
plot the limits on lighter chargino and lighter stop mass for chosen
$M_A$ and $M_2/\mu$ values. In both plots we scan over $\theta_{LR}$
in the range $-60^{\circ}<\theta_{LR}<60^{\circ}$.

	Fig.~\ref{fig:bsg_ma} shows that if $M_2/|\mu|$ is small, so
that lighter chargino consists predominantly of gaugino and its
contribution to \BSG is small, limits on $M_A$ are quite strong. e.g.
$M_A\geq{\cal O}(200~\mathrm{GeV})$ for
$m_{\tilde{T_1}}=m_{\chi_1^{\pm}}=90$ GeV (we take 95\% errors of CLEO
measurement). The limits decrease when $M_2/|\mu|$ increases and
approximately saturate for $M_2/|\mu|\geq 1$. Similar effects are
visible in Fig.~\ref{fig:bsg_ct} where {\em upper} bounds on
$(m_{\tilde{T_1}}$ and $m_{\chi_1^{\pm}})$ are shown. For small
$M_2/|\mu|$, very light stop and chargino are necessary to cancel the
charged Higgs contribution. Thus, the corresponding upper limits on
their masses are very strong. For large $M_2/|\mu|$, chargino and stop
even 2-3 times heavier than the charged Higgs are allowed.

\section{Implications of the FCNC bounds for SUSY breaking and
sfermion mass generation}
\label{sec:rge}

	We have already said in section~\ref{sec:ckm} that the problem
of achieving the necessary suppression of SUSY contributions to FCNC
processes admits two broad classes of solutions (besides
fine-tuning). One of them relies on the possibility that sfermions of
the first two generations are much heavier than the usually expected
scale of supersymmetry breaking ${\cal O}(1~\TeV)$. In this case, the
suppression of FCNC effects in the MSSM is achieved with essentially
arbitrary flavour structure of squark mass matrices. The other
possibility is that sfermion mass matrices at $M_Z$ energy scale are
indeed, for some reason, almost flavour conserving in the super-KM
basis, and have approximately degenerate diagonal mass terms of the
first two generation sfermions.  Here, we would like to discuss and
summarize various theoretical aspects and ideas behind these two
general approaches to the FCNC problem in the MSSM.

Beginning with the possibility of heavy sfermions of the first two
generations, one should stress (see e.g.~\cite{HIER12}) that it is
consistent with supersymmetry remaining the solution to the hierarchy
problem. Indeed, the minimization of the Higgs potential in the MSSM
gives
\bea
M_Z^2 = {2\over \tan^2\beta - 1}\left(M_{H^1}^2 + |\mu|^2 
- (M_{H^2}^2+ |\mu|^2)\tan^2\beta\right)
\label{eq:vmin}
\eea
where $M_{H^1}$ and $M_{H^2}$ are the soft Higgs masses defined in
eq.~(\ref{eq:lsoft}) and $\mu$ is defined in eq.~(\ref{eq:superpot}).
The hierarchy problem is avoided so long as we do not introduce large
cancellations in eq.~(\ref{eq:vmin}), after expressing $M_{H^1}$ and
$M_{H^2}$ in terms of the soft masses at $M_{GUT}$. Using the
appropriate RG equations one obtains
\bea
M_Z^2 &\simeq & 
  a_{\scs H^1} M^2_{H^1}(M_{\scs GUT})
+ a_{\scs H^2} M^2_{H^2}(M_{\scs GUT}) 
+ a_{\scs QU} [ (M_Q^2)^{33}(M_{\scs GUT}) + (M_U^2)^{33}(M_{\scs GUT})
\vspace{-0.2cm}
\nonumber \eea
\bea
-2 |\mu(M_Z)|^2 
+ a_{\scs AA} [A_U^{33}(M_{\scs GUT})]^2
+ a_{\scs AM} A_U^{33}(M_{\scs GUT}) M_{1/2}(M_{\scs GUT})
+ a_{\scs MM} M_{1/2}^2(M_{\scs GUT}).~~~~~&&
\label{eq:rgemass}
\eea
For $m_t = 175$~GeV and $\tan\beta = 1.65(2.2)$, the values of the
coefficients are the following \cite{CP97}: 
$a_{\scs H^1} = 1.2(0.5)$,
$a_{\scs H^2} = 1.7(1.5)$,
$a_{\scs QU} = 1.5(1.1)$,
$a_{\scs AA} = 0.1(0.2)$,
$a_{\scs AM} = -0.3(-0.7)$ and
$a_{\scs MM} = 15.0(10.8)$.
The crucial observation is that to a very good approximation (see
ref.~\cite{HIER12} for corrections), the first two generation sfermion
masses do not enter into the above expression for $M_Z$. The dominant
role in this expression is played by the common gaugino mass $M_{1/2}$
at $M_{GUT}$ and the third generation sfermion masses. The constraint
on the first two generations (via $D$-term in the Lagrangian) is very
weak and admits masses $m_{1,2} \sim {\cal O}(10~\TeV)$. Thus, large
non-universality with $m_{1,2}\sim{\cal O}(10~\TeV)$ and $m_3\leq
1~\mathrm{TeV}$ is consistent with the absence of large cancellations
in the Higgs potential. On the other hand, we observe that $M_{1/2}$
and $m_3$ have to be rather small at $M_{GUT}$. In particular, large
coefficient in eq.~(\ref{eq:rgemass}) requires very light chargino.
The low energy stop mass parameters are actually even smaller than
$m_3$ at $M_{GUT}$ due to the running with large $Y_t$. Thus, the
scenario with light chargino and stop is not unattractive. The
discussed solution does not require any constraints on flavour
off-diagonal entries in the (1,2) sector.
 
	Another way to solve the FCNC problem in the MSSM is to embed this
model into a high energy theory which assures small $\delta^{IJ}$ at
the $M_Z$ scale. Several ideas along these lines have been discussed
in the literature. Generally speaking, this option correlates much
stronger the FCNC problem with the theory of soft SUSY breaking and/or
fermion mass generation.

	Before going into further details we would like to address one
interesting renormalization effect: Very small $\delta^{IJ}$ at $M_Z$
do not necessarily imply that $\delta^{IJ}$ are small at
$M_{GUT}$. Indeed, as it has been shown in ref.~\cite{CEKLP95}, there
are QCD renormalization effects which increase only diagonal entries
in the sfermion mass matrices. Consequently, they suppress
$\delta^{IJ}$. These effects can become dramatic when scalar masses at
$M_{GUT}$ are much smaller than $M_{1/2}$. Then even $\delta^{IJ} \sim
{\cal O}(1)$ is acceptable in the squark sector at $M_{GUT}$. This
could still happen with gaugino masses being ${\cal O}(M_Z)$. There
are no absolute lower bounds on scalar masses at $M_{GUT}$. If they
are much smaller than $M_{1/2}$, then this latter parameter sets the
magnitude of physical masses for basically all the superpartners.

	Such a scenario is not free of problems either. So long as no
theoretical reason is found for $M_{1/2}$ being much larger than other
soft SUSY breaking parameters at $M_{GUT}$, this has to be understood
as certain fine-tuning. Moreover, suppression of $\delta^{IJ}$ in the
lepton sector is much less efficient because electroweak
renormalization effects are smaller. One may conclude that although
renormalization effects are helpful, they do not eliminate the flavour
problem. If $\delta^{IJ}$ is small at the $M_Z$ scale, we need a
theory of flavour which assures it.

	A simpler way to account for small $\delta^{IJ}$ at $M_Z$ is to
assure that the soft supersymmetry breaking scalar masses are
generated as flavour diagonal and degenerate, and that the trilinear
$A$-terms are proportional to the Yukawa couplings with a universal
mass coefficient~\cite{SQDEGEN}. Thus, the absence of strong FCNC
effects in SUSY would be explained by a particularly simple pattern of
soft supersymmetry breaking (``universal soft terms''), with no
correlation to the fermion mass generation. This scenario can be
obtained under the assumption of dilaton dominance in the supergravity
models for soft supersymmetry breaking \cite{SP4}. However, deeper
understanding of neither such a dominance nor the stabilization of the
dilaton potential is available yet. Another possibility is gauge
mediated supersymmetry breaking at low energies \cite{SP1}, which
naturally leads to almost universal soft terms.

	Quark-squark mass alignment is a different idea which relies on
strong correlation between fermion and sfermion mass generation. Many
different models of that type have been proposed. The most popular
ones explore horizontal symmetries or an anomalous $U(1)$
symmetry~\cite{SP2}.

	Both options, if realized in an exact way, leave no room for
flavour violation in the sfermion mass matrices, up to small ${\cal
O}(K^{IJ})$ renormalization effects. However, in most ``realistic''
models of both types, there are interesting departures from the exact
realization. For instance, in the supergravity models with Grand
Unified groups, it is natural to assume universal soft terms at the
Planck scale rather than at the GUT scale. The RG running down to the
GUT scale generates flavour mixing in both the squark and slepton mass
matrices at $M_{GUT}$. This gives interesting effects for $l^I\ra
l^J\gamma$~\cite{SP3}, within the reach of the forthcoming
experiments. Moreover, the evolution down to low energies gives
generically light stop, i.e. the scenario we have discussed in
section~\ref{sec:ckm}.

	Similar ``inverse hierarchy'' of sfermion masses (with respect to
fermion ones) is also obtained in models with a $U(1)$
symmetry~\cite{INVH}. The quark-squark mass alignment is generically
not perfect, and the FCNC effects are expected to be not much below
the present experimental bounds.

\section{Summary}
\label{sec:summary}

	There exist important bounds on new sources of FCNC in the
MSSM. When the SUSY breaking scale is around 1~TeV, some of the
flavour off-diagonal entries in the squark mass matrices have to be an
order of magnitude lower than the diagonal ones. Most severe
constraints exist in the left-right squark mixing sector. However,
they could be considered dramatic only when flavour-conserving
left-right mixing was generically large. Bounds on the squark mass
degeneracy are in the range of few tens of percent even for the first
two generations, so long as their masses are around 1~TeV. Thus, we
conclude that the supersymmetric flavour problem is intriguing but
perhaps not as severe as it has been commonly believed.

	Restrictions on supersymmetric flavour violation have no
immediate explanation in terms of the basic structure of the MSSM.
They provide an important hint on its embedding into a more
fundamental theory of soft supersymmetry breaking and fermion mass
generation. Interesting effects in \kk, \bb and particularly in \bsg
can be expected from light stop and chargino. Interesting effects can
be expected in the lepton sector $(l^I\ra l^J\gamma)$, as well. Models
of soft supersymmetry breaking which are consistent with the existing
bounds leave room for such new effects. New sector of ``precision
experiments'' ({\'a} la LEP) is welcome!

\renewcommand{\thesection}{Note Added}
\section{}

	The expressions for chargino contributions to \bsg presented
in our appendix allow to derive a bound on $(\delta_U^{23})_{LR}$.
Requiring that the second term in eq.~(\ref{eq:apmsum}) gives smaller
contribution to the amplitude than the Standard Model, we find
\be \label{LRbound}
|(\delta_U^{23})_{LR}| \; \simleq 12 \; \left( \f{m_0}{1~{\rm TeV}} \right)^2 \; F,
\ee
where $m_0$ is the average up-squark mass, and the factor $F$ is of
order unity or larger. The explicit expression for $F$ is
\be
F = \; sin \beta \; 
\left|       x_1^2 f'_1(x_1) Z^{+*}_{11} Z^+_{21}
       \;+\; x_2^2 f'_1(x_2) Z^{+*}_{12} Z^+_{22} \right|^{-1},
\ee
where $x_p = m^2_0/m^2_{\chi_p^{\pm}}$ and $f'_1(x)$ is the derivative
of the function $f_1$ given below eq.~(\ref{eq:c7ch}). Even when $F =
1$, the bound on the r.h.s. of eq.~(\ref{LRbound}) is smaller than
unity (i.e. effective) only for the average squark mass below around
300~GeV.

	The same bound holds for $(\delta_U^{13})_{LR}$, because we
know from experiment that $b \ra d \gamma$ does not have significantly
larger rate than the SM prediction for \bsg. Bounds on
$(\delta_U^{13})_{LR}$ and $(\delta_U^{23})_{LR}$ can be derived from
chargino contributions to \bb mixing, too.

	We thank Luca Silvestrini and Andrea Romanino for bringing
bounds on $(\delta_U^{13})_{LR}$ and $(\delta_U^{23})_{LR}$ to our
attention.

\renewcommand{\thesection}{Acknowledgements}
\section{}

	We thank Piotr Chankowski, Antonio Masiero and Luca
Silvestrini for helpful discussions.

	This work was supported in part by the Polish Commitee for
Scientific Research under grants 2~P03B~040~12~(1997-98) and
2~P03B~180~09~(1995-97) and by EC contract HCMP CT92004. M.M. was
supported in part by Schweizerischer Nationalfonds. J.R. was supported
in part by Alexander von Humboldt Stiftung.

\renewcommand{\thesection}{Appendix~\Alph{section}}
\renewcommand{\thesubsection}{\Alph{section}.\arabic{subsection}}
\renewcommand{\theequation}{\Alph{section}.\arabic{equation}}

\setcounter{equation}{0}
\setcounter{section}{0}

\section{}
\label{app:bsgct}

 	As an example, we discuss in more detail the chargino--(up~squark)
contribution to \bsg decay rate. This contribution is proportional to
the quantity $C_7^{(0)\chi^{\pm}}(M_W)$ given in eq.~(\ref{eq:c7ch}).
In order to obtain the expression for $C_7^{(0)\chi^{\pm}}(M_W)$ in
the mass insertion approximation discussed in Sec.~\ref{sec:ndiag}, we
expand the exact formulae~(\ref{eq:c7ch}) up to the first order in the
physical up-squark mass splitting:
\bea
m^2_{\tilde{U}_i}& =& m^2_0 + \delta m^2_{\tilde{U}_i}
\eea
The functions $f_1$ and $f_2$ can be expanded around the average mass
$m_0^2$ as
\bea
f_1\left({m^2_{\tilde{U}_i}\over m^2_{\chi_p^{\pm}}}\right)
& \approx & 
f_1\left({m^2_0\over m^2_{\chi_p^{\pm}}}\right) +
{m^2_{\tilde{U}_i}-m^2_0\over  m^2_{\chi_p^{\pm}}}
f_1'\left({m^2_0\over m^2_{\chi_p^{\pm}}}\right)
\nonumber\\
&=&\alpha_1^p + \beta_1^p  m^2_{\tilde{U}_i}\;,
\eea
where
\bea
\alpha_1^p &=&  f_1\left({m^2_0\over m^2_{\chi_p^{\pm}}}\right) -
{m^2_0\over m^2_{\chi_p^{\pm}}} 
f_1'\left({m^2_0\over m^2_{\chi_p^{\pm}}}\right) \\
 \beta_1^p&=&{1\over m^2_{\chi_p^{\pm}}}
f_1'\left({m^2_0\over m^2_{\chi_p^{\pm}}}\right),
\eea
and similarly for the function $f_2$.  After such expansion,
eq.~(\ref{eq:c7ch}) may be rewritten in the form
\bea
C_7^{(0)\chi^{\pm}}(M_W) &=& 
\f{1}{K_{ts}^* K_{tb}} \sum_{p=1}^2 \f{M_W^2}{m_{\chi_p^{\pm}}^2} 
\left\{ 
-\f{1}{6}\alpha^p_1 \sum_{i=1}^6  A^{p*}_{i2} A^p_{i3} 
-\f{1}{6}\beta^p_1 \sum_{i=1}^6  A^{p*}_{i2} A^p_{i3} m^2_{\tilde{U}_i}
\right.\nonumber\\
&+&\left. \f{m^2_{\chi_p^{\pm}}}{3 m_b} \left[
\alpha^p_2 \sum_{i=1}^6 A^{p*}_{i2} B^p_{i3} 
+ \beta^p_2 \sum_{i=1}^6 A^{p*}_{i2} B^p_{i3} m^2_{\tilde{U}_i} \right]
\right\} \; + \; {\cal O}((\delta m^2/m_0^2)^2). \hspace{1.5cm}
\label{eq:c7che}
\eea
Sums over $i$ in the eq.~(\ref{eq:c7che}) can be calculated using
unitarity of the matrix $Z_U$ and eq.~(\ref{eq:zudef})
\bea
\sum_{k=1}^6 Z_U^{ik}Z_U^{jk\star} &=& \hat{\mbox{\large 1}}^{ij}\\
\sum_{k=1}^6 Z_U^{ik}Z_U^{jk\star}m^2_{\tilde{U}_i} &=&
\left({\cal M}^2_{\tilde{U}}\right)^{ij}
\eea
Evaluating the four sums appearing in eq.~(\ref{eq:c7che}) gives the
following results:
\bea
\sum_{i=1}^6 A^{p\star}_{i2}A^p_{i3}& = & 
{1\over 2M_W^2 \sin^2\beta} |Z^+_{2p}|^2 
\left[K^{\dagger}m_u^2 K\right]_{23}
\label{eq:apsum}
\\
\sum_{i=1}^6 A^{p\star}_{i2}B^p_{i3}& = & 0\label{eq:bpsum}
\\
\sum_{i=1}^6 A^{p\star}_{i2}A^p_{i3}m^2_{\tilde{U}_i}& = &
|Z^+_{1p}|^2  \left[K^{\dagger}(M^2_{\tilde{U}})_{LL} K 
+ K^{\dagger}m_u^2  K\right]_{23}\nonumber\\ 
&-& {1\over \sqrt{2} M_W \sin\beta}Z^{+\star}_{1p}Z^+_{2p}
\left[K^{\dagger}(M^2_{\tilde{U}})_{LR}m_u K
- \mu\cot\beta K^{\dagger}m_u^2 K\right]_{23}\nonumber\\
&-& {1 \over \sqrt{2} M_W \sin\beta}Z^{+}_{1p}Z^{+\star}_{2p}
\left[K^{\dagger}m_u (M^2_{\tilde{U}})^{\dagger}_{LR} K
- \mu^{\star}\cot\beta K^{\dagger}m_u^2 K\right]_{23}\nonumber\\
&+& {1\over 2 M_W^2 \sin^2\beta}|Z^+_{2p}|^2 
\left[K^{\dagger}m_u(M^2_{\tilde{U}})_{RR}m_u K 
+ K^{\dagger}m_u^4  K\right]_{23}\label{eq:apmsum}
\\
\sum_{i=1}^6 A^{p\star}_{i2}B^p_{i3}m^2_{\tilde{U}_i}& = &
-{m_b\over \sqrt{2}M_W\cos\beta} Z^{+\star}_{1p} Z^{-\star}_{2p}
\left[K^{\dagger}(M^2_{\tilde{U}})_{LL}K 
+ K^{\dagger}m_u^2  K\right]_{23}\nonumber\\
&+& {m_b\over M_W^2\sin2\beta} Z^{+\star}_{2p} Z^{-\star}_{2p}
\left[K^{\dagger}m_u(M^2_{\tilde{U}})^{\dagger}_{LR} K
- \mu^{\star}\cot\beta K^{\dagger}m_u^2 K\right]_{23}\label{eq:bpmsum}
\nonumber\\
\eea
Inserting the above sums into the eq.~(\ref{eq:c7che}), we obtain an
approximate expression for the chargino contribution to \bsg decay
amplitude expressed in terms of the initial squark mass matrices.

	As seen in eqs.~(\ref{eq:apsum})--(\ref{eq:bpmsum}), there are
many terms in this contribution which survive in the limit of
flavour-conserving squark mass matrices. The FCNC effects we have
discussed in section~\ref{sec:ckm} originate from these terms.

	An important observation is that, as follows from
eq.~(\ref{eq:udcorr}), $(M^2_{\tilde{U}})_{LL} = K
(M^2_{\tilde{D}})_{LL} K^{\dagger}$. After inserting this relation
into eqs.~(\ref{eq:apsum})--(\ref{eq:bpmsum}), we come to an immediate
conclusion that, as far as the left squarks are concerned, chargino
contributions to \bsg are directly sensitive to the structure of down
(not up!) left squark mass matrix (as the diagrams with gluino and
neutralino exchanges are).  The situation with right squark mass
matrices is more complicated: In the expressions for chargino
contributions, they are multiplied by quark masses and the KM
matrices, which causes that the final result is sensitive to the
linear combination of various off-diagonal entries and also to the
splitting of the diagonal elements. However, most of these terms are
suppressed by light quark masses.\footnote{
The only exception is the second term in eq.~(\ref{eq:apmsum}) which
gives a bound on $(\delta_U^{23})_{LR}$.}
For instance, $K^{\dagger}m_u(M^2_{\tilde{U}})_{RR}m_u K$ can be
written as
\bea 
K^{\dagger}m_u(M^2_{\tilde{U}})_{RR}m_u K &\approx &
m_t^2(M^2_{\tilde{U}})_{RR}^{33}\left(\begin{array}{ccc}
0&0&K^{31\star}\\
0&0&K^{32\star} + \frac{m_c}{m_t}
\frac{(M^2_{\tilde{U}})_{RR}^{23}}{(M^2_{\tilde{U}})_{RR}^{33}}\\
K^{31}& K^{32} + \frac{m_c}{m_t}
\frac{(M^2_{\tilde{U}})_{RR}^{32}}{(M^2_{\tilde{U}})_{RR}^{33}}&
1\\
\end{array}\right)\nonumber\\
& +& \mathrm{smaller~terms}
\eea
Therefore $\left(K^{\dagger}m_u(M^2_{\tilde{U}})_{RR}m_u K\right)_{23}
\approx m_t^2 (M^2_{\tilde{U}})_{RR}^{33}K^{32\star} + m_c m_t
(M^2_{\tilde{U}})_{RR}^{23}$ is basically sensitive only to the
elements $(M^2_{\tilde{U}})_{RR}^{23},
(M^2_{\tilde{U}})_{RR}^{33}$.

	Another (and most obvious) simplification of
eqs.~(\ref{eq:apsum}-\ref{eq:bpmsum}) is achieved by neglecting light
quark masses in
\bea
(K^{\dagger}m_u^n K)_{23}\approx K^{32\star} K^{33} m_t^n 
= K^{\star}_{ts} K_{tb} m_t^n,~~~~~~~~~n=2,4.
\eea

	The observation that chargino contributions to the \bsg depend on
the down, not up, left squark mass matrix may be generalized to other
processes. One always finds that chargino, neutralino and gluino
contributions to the processes involving down quarks in the initial
and final states are sensitive to the structure of the left down
squark matrices only. It follows from the structure of the up-squark
mass matrix (eq.~(\ref{eq:sfmass})) and $u\tilde{D}\chi$ vertex shown
in Fig.~\ref{fig:ccurrent}. One can verify this by writing the matrix
$Z_U$ in the form
\bea
Z_U = 
\left(\begin{array}{cc}
K&0\\
0&1\\
\end{array}\right)X_U
\eea
The new ``rotated'' diagonalization matrix $X_U$ is defined by the
following condition:
\bea 
\left({\cal M}^2_{\tilde{U}}\right)^{diag} = 
X_U^{\dagger} {\cal M}^2_{X\tilde{U}} X_U
\label{eq:xudef}
\eea
where ${\cal M}^2_{X\tilde{U}}$ is explicitly dependent on the left
down squark mass matrix $(M^2_{\tilde{D}})_{LL}$
\bea 
{\cal M}^2_{X\tilde{U}} = 
\left( \begin{array}{cc}
(M^2_{\tilde{D}})_{LL} + K^{\dagger}m_u^2 K
- \frac{\cos 2\beta}{6}(M_Z^2 - 4M_W^2)\hat{\mbox{\large 1}} &
K^{\dagger}\left((M^2_{\tilde{U}})_{LR} 
- \cot\beta \mu m_u\right)\\
\left((M^2_{\tilde{U}})_{LR}^{\dagger} 
- \cot\beta \mu^{\star}  m_u\right)K& 
(M^2_{\tilde{U}})_{RR} + m_u^2
+\frac{2\cos 2\beta}{3} M_Z^2\sin^2\theta_W\hat{\mbox{\large 1}} \\ 
\end{array}\right)\nonumber\\
\label{eq:uxmass}
\eea
The $u\tilde{D}\chi$ vertex written in terms of $X_U$ takes the form
\bea
i\left[\left(-g_2
X_{U}^{Ji*} Z_{1j}^+ 
+ K^{JI}Y_u^{J} X_{U}^{(J+3)i*} Z_{2j}^+\right) P_L 
- Y_d^{I} X_{U}^{Ji*} Z_{2j}^{-*} P_R \right]
\label{eq:xuds}
\eea
In this expression, the KM matrix occurs only multiplied by the
up-quark Yukawa couplings which originate from the
higgsino--(right~squark) interactions.

	Similarly, one can prove that processes involving up quarks in the
initial and final states (like \dd mixing) are sensitive only to the
structure of the left up squark mass matrix.

\end{document}